\documentclass[a4paper,11pt]{article}
\pdfoutput=1
\usepackage[utf8x]{inputenc}
\usepackage{jheppub}
\usepackage{epsf}
\usepackage{graphicx}	% needed for figures
\usepackage{dcolumn}	% needed for some tables
\usepackage{bm}			% for math
\usepackage{amssymb,amsmath}
\usepackage{bm}
\usepackage{wasysym}
\usepackage{lipsum, color}
\usepackage{physics}
\usepackage{bbm}		% \mathbbm{1} - the unit matrix.
\usepackage{multirow}
\usepackage{slashed}
\usepackage{hyperref}
\usepackage[usenames,dvipsnames,svgnames]{xcolor}
\definecolor{rossos}{cmyk}{0,1,1,0.55}
\definecolor{bluscuro}{rgb}{0.15, 0.2, .85}
\definecolor{bluchiaro}{cmyk}{1,.3,0.,0.1}
\hypersetup{colorlinks, citecolor=bluscuro, linkcolor=bluscuro, urlcolor=bluscuro}

\let\oldquote\quote
\renewcommand\quote{\scriptsize\oldquote}
\let\oldquotation\quotation
\renewcommand\quotation{\scriptsize\oldquotation}

\newcommand{\gsim}{\lower.7ex\hbox{$\;\stackrel{\textstyle>}{\sim}\;$}}
\newcommand{\lsim}{\lower.7ex\hbox{$\;\stackrel{\textstyle<}{\sim}\;$}}

\def\beq{\begin{equation}}
\def\eeq{\end{equation}}
\def\be{\begin{equation}}
\def\ee{\end{equation}}
\def\bea{\begin{eqnarray}}
\def\eea{\end{eqnarray}}
\def\bmat{\begin{pmatrix}}
	\def\emat{\end{pmatrix}}
\def\bei{\begin{itemize}}
	\def\eei{\end{itemize}}

\def\hpi{\hat{\pi}}
\def\hK{\hat{K}}
\def\he{\hat{\eta}}

\usepackage[normalem]{ulem}

\def\ie{{\it {i.e.}},~}
\def\eg{{\it {e.g.}},~}

\makeatletter
\def\section{\@startsection {section}{1}{\z@}{-3.5ex plus -1ex minus
		-.2ex}{2.3ex plus .2ex}{\large\bf}}
\def\subsection{\@startsection{subsection}{2}{\z@}{-3.25ex plus -1ex
		minus -.2ex}{1.5ex plus .2ex}{\normalsize\bf}}
\makeatother
\makeatletter

\@addtoreset{equation}{section}

\makeatother

\usepackage{graphicx}  % needed for figures
\usepackage{dcolumn}   % needed for some tables
\usepackage{bm,relsize}        % for math
\usepackage{amssymb, amsmath}
\usepackage{wasysym}
\usepackage{slashed}
\usepackage{bbold} % For Aviv
\usepackage{mathtools}
\usepackage{comment}
\usepackage{cancel}
\usepackage{ulem}

\usepackage{feyn}
\def\beq{\begin{equation}}
\def\eeq{\end{equation}}

\begin{document}

\global\long\def\com#1#2{\underset{{\scriptstyle #2}}{\underbrace{#1}}}

\global\long\def\comtop#1#2{\overset{{\scriptstyle #2}}{\overbrace{#1}}}

\global\long\def\ket#1{\left|#1\right\rangle }

\global\long\def\bra#1{\left\langle #1\right|}

\global\long\def\braket#1#2{\left\langle #1|#2\right\rangle }

\global\long\def\op#1#2{\left|#1\right\rangle \left\langle #2\right|}

\global\long\def\opk#1#2#3{\left\langle #1|#2|#3\right\rangle }

\global\long\def\L{\mathcal{L}}

\title{Dark Matter on a Slide}

\author[a]{Hsin-Chia Cheng,}
\author[b,c,d]{Xu-Hui Jiang,}
\author[d,e,f]{Lingfeng Li,}
\author[g,h]{Ennio Salvioni,$\,$}
\affiliation[a]{Center for Quantum Mathematics and Physics (QMAP), Department of Physics,\\
University of California, Davis, CA 95616, USA}
\affiliation[b]{Center for Future High Energy Physics, Institute of High Energy Physics,\\ Chinese Academy of Sciences, Beijing, China}
\affiliation[c]{China Center of Advanced Science and Technology, Beijing, China}
\affiliation[d]{International Center of Theoretical Physics-Asia Pacific,\\ University of Chinese Academy of Sciences, Beijing 100190, China}
\affiliation[e]{Institute of High Energy Physics, Beijing 100049, China}
\affiliation[f]{Department of Physics, Brown University, Providence, RI 02912, USA}
\affiliation[g]{Departament de F\'isica, Universitat Aut\`onoma de Barcelona, 08193 Bellaterra, Barcelona, Spain}
\affiliation[h]{IFAE and BIST, Campus UAB, 08193 Bellaterra, Barcelona, Spain}

\date{\today}

\emailAdd{hcheng@ucdavis.edu, jiangxuhui@ucas.ac.cn, l.f.li165@gmail.com, esalvioni@ifae.es}

\abstract{We present a scenario for GeV-scale thermal dark matter that can only be tested with accelerator experiments. Dark matter is composed of dark pions arising from a confining strong interaction in the dark sector. The thermal relic density is obtained through the interplay of up-scatterings of dark pions to heavier dark mesons (the dark counterparts of the kaons and $\eta$), and decays of the unstable dark $\eta$ to Standard Model particles. This mechanism is analogous to a playground slide, where one climbs up first and then slides down with a release of energy. We illustrate the scenario with a minimal model based on the $SU(3)/SO(3)$ coset, where dark matter is stabilized by a $U(1)$ flavor symmetry. The correct relic density is obtained with dark meson mass splittings of $10\%$ to $50\%$ and a dark-$\eta$ lifetime shorter than $10^3\,\mathrm{m}/c$. Direct and indirect dark matter searches are mostly ineffective, as a consequence of the charge conjugation symmetry of the stabilizing $U(1)$. The most striking signals arise at the LHC, from the production of dark showers containing long-lived dark $\eta$’s that decay to visible final states. These signatures crucially depend on the portal interaction connecting the dark sector to the Standard Model. We show that several well-known portals can complete the scenario above the weak scale, and outline the expected signals in each case.

}

\maketitle

\section{Introduction}\label{sec:intro}

A strongly-interacting dark sector~\cite{Strassler:2006im} can naturally host dark matter (DM) in the form of a bound state of dark quarks. While dark baryons typically require an asymmetry between dark baryons and antibaryons to obtain the observed DM relic density, dark mesons can do so through the freeze-out of thermal processes. A well-known example is the strongly interacting massive particle (SIMP) paradigm~\cite{Hochberg:2014dra,Hochberg:2014kqa}, where pion-like dark mesons undergo $3\to 2$ annihilations mediated by Wess-Zumino-Witten (WZW) terms~\cite{Wess:1971yu,Witten:1983tw,Witten:1983tx}. The freeze-out of $3\to 2$ annihilations selects the GeV scale for the DM mass.

In this work, we explore another generic mechanism that allows GeV-scale dark pions to acquire a thermal abundance. The dark pions forming DM are accompanied by other, heavier dark mesons, some of which are unstable and decay back to Standard Model (SM) particles. The DM relic density is determined by either the freeze-out of forbidden annihilations~\cite{Griest:1990kh,DAgnolo:2015ujb} to the heavier dark mesons, or the decoupling of decays of the unstable species to the SM, depending on which happens earlier. Because of the analogy to a playground slide, where one needs to climb up the ladder first, then slide (decay) down, we call this mechanism {\it slide dark matter.}  For moderate mass splittings, in the tens of percent, the mechanism selects GeV-scale DM masses. 

Slide DM can arise naturally in a (three-flavor) dark copy of QCD: if dark isospin is an exact symmetry, then the pions and the heavier kaons are stable, while the heaviest $\eta$ meson can decay to SM particles, as it is a singlet under isospin. DM consists mostly of dark pions, with a small fraction made of dark kaons.\footnote{With two dark flavors, the unstable neutral pion is expected to be lighter than the charged pions, so the mechanism cannot be at play.} The possibility of dark pion DM accompanied by other unstable dark pions was studied before~\cite{Buckley:2012ky,Kopp:2016yji,Beauchesne:2018myj,Beauchesne:2019ato,Carmona:2024tkg}, and the QCD-like setup described above was already classified in Ref.~\cite{Beauchesne:2019ato}. However, those references considered larger masses, typically of order $100\;\mathrm{GeV}$, and/or spectra where the stable DM mesons and the unstable species are (almost) mass-degenerate, with the DM relic density being determined by the co-decaying mechanism~\cite{Dror:2016rxc,Farina:2016llk}. Here we focus on GeV-scale dark mesons with sizeable mass splittings.

To present the slide DM mechanism, we introduce a more minimal model. We consider the coset $SU(3)/SO(3)$, containing $5$ pseudo-Nambu-Goldstone-boson (pNGB) dark mesons. The stability of DM is protected by a $U(1)$ flavor symmetry (instead of the $SU(2)$ isospin symmetry of the QCD-like example). From lightest to heaviest, the pNGBs are the dark pions $\hpi$ (with charges $\pm 1$ under the $U(1)$), the dark kaons $\hK$ (charges $\pm 1/2$) and the dark eta $\he$ (neutral). Their masses are split due to the different masses of the dark quarks. The DM relic density depends on the pNGB masses and splittings, their self-interaction strength $1/f_{\hpi}$, where $f_{\hpi}$ is the dark analogue of the pion decay constant, as well as the $\he$ decay width. As the DM relic density depends on the decay width, it is insensitive to the nature of the portal interaction mediating $\he$ decays to SM particles.

A remarkable feature of this model is that vector-current interactions between DM and the SM fields are forbidden by the charge conjugation symmetry of the stabilizing $U(1)$. As a result, both direct and indirect searches for DM are ineffective in most of the parameter space. The most promising experimental probes arise from the production of dark hadrons at accelerator experiments, chief among them the LHC. There, the nature of the portal interaction plays a crucial role in devising search and analysis strategies.

In this work, we show that several well-known classes of portal interactions can complete the slide DM mechanism above the weak scale, and we discuss the expected experimental signatures in each case. We find that the decay length of the unstable dark meson $\he$ ranges from sub-millimeter to kilometer in most of the viable parameter space. Consequently, striking dark shower signals are predicted at the LHC~\cite{Albouy:2022cin}, which could manifest as semi-visible jets~\cite{Cohen:2015toa}, emerging jets~\cite{Schwaller:2015gea}, displaced decays of long-lived particles (LLPs)~\cite{Strassler:2006im,Strassler:2006qa,Han:2007ae,Alimena:2019zri}, non-standard jet substructure~\cite{Park:2017rfb,Cohen:2023mya}, or their combinations. 

A related analysis of dark showers in a model containing stable and unstable dark mesons appeared in Ref.~\cite{Carmona:2024tkg}, where indirect detection constraints were alleviated by assuming exact mass degeneracy of all the dark mesons, while direct detection bounds remained significant. The model we present naturally avoids all constraints from direct and indirect DM searches, due to the mass splittings among the pNGBs and the dark charge conjugation symmetry. In Ref.~\cite{Bernreuther:2019pfb} the unstable species was formed by vector resonances, which requires proximity of the pNGBs and the vector mesons; here we focus on the small dark quark mass regime, where the vector mesons are negligible.

The rest of the paper is organized as follows. In Sec.~\ref{sec:model}, we introduce the $SU(3)/SO(3)$ model and discuss the mass spectrum and interactions of the pNGB dark mesons. The thermal evolution and DM relic density are analyzed in Sec.~\ref{sec:thermal}, where the viable parameter space is also presented. In Sec.~\ref{sec:ID} we show that, although direct and indirect DM searches are mostly ineffective in this model, an exception can occur for small mass splittings: if the $\hK$'s constitute a significant fraction of DM, their annihilation may lead to indirect detection signals. In Sec.~\ref{sec:UV} we present three classes of ultraviolet-complete portal interactions that can mediate the decays of $\he$ to SM particles, and outline the expected signals at the LHC: in Sec.~\ref{sec:Z-portal}, the $Z$-portal scenario, where heavy dark quarks with SM electroweak charges are introduced~\cite{Cheng:2019yai,Cheng:2021kjg}; in Sec.~\ref{sec:zprime}, a heavy $Z'$ mediator~\cite{ParticleDataGroup:2024cfk} with chiral couplings to the SM fermions; and in Sec.~\ref{sec:bifund}, heavy scalars transforming under both the SM color and dark color gauge symmetries~\cite{Bai:2013xga,Schwaller:2015gea,Beauchesne:2017yhh,Renner:2018fhh,Carmona:2024tkg}. Conclusions are drawn in Sec.~\ref{sec:conclusions}. Appendix~\ref{app:Boltzmann} contains the complete Boltzmann equations for the $SU(3)/SO(3)$ model. The effects of scalar-current interactions between the dark pions and the SM, which are induced by the Higgs field in the $Z$-portal completion, are discussed in Appendix~\ref{app:Higgs}. Appendix~\ref{app:phenomenology} provides further details on the LHC sensitivity projections presented in Sec.~\ref{sec:Z-portal}.

\section{The $SU(3)/SO(3)$ Model}
\label{sec:model}
We consider a non-Abelian $SO(N_d)$ dark gauge group, with $N_f = 3$ light flavors of Weyl dark quarks transforming in the fundamental representation. The dark quarks are denoted as $\psi_i,\, i=1,2,3$ and assumed to be left-handed. We focus on $N_d \geq 4$, where QCD-like chiral symmetry breaking is expected~\cite{Lee:2020ihn}; for $N_d = 3$, lattice results indicate that the theory lies inside the conformal window~\cite{Bergner:2017gzw}. At low energies, as the gauge coupling becomes strong, the dark quarks form a condensate $\langle \psi_i \psi_j \rangle = \mu^3 \delta_{ij}$, breaking spontaneously the $SU(N_f)$ flavor symmetry $\bm{\psi}\to U \bm{\psi}$ down to $SO(N_f)$ and yielding $\tfrac{1}{2}N_f (N_f + 1) - 1 = 5$ pNGBs in the low-energy spectrum. The pNGBs transform in the two-index traceless symmetric tensor representation of $SO(N_f)$. The unbroken generators satisfy $T_a + T_{a}^{T} = 0$ ($a = 2, 5, 7$) and the broken ones $X_i - X_{i}^T = 0$ ($i = 1, 3, 4, 6, 8$), where $ T_a\, (X_i)  = \lambda_{a (i)} / 2$, with $\lambda_{a (i)}$ denoting the standard Gell-Mann matrices. Since the dark quarks lie in a real representation of $SO(N_d)$, and are assumed neutral under the SM gauge group, they admit Majorana mass terms 
\begin{equation}\label{eq:L_psi}
\mathcal{L}_\psi =  \bm{\psi}^\dagger i \bar{\sigma}^\mu D_\mu \bm{\psi}  -\, \frac{1}{2} \big( \bm{\psi}^T \mathcal{M} \epsilon\hspace{0.2mm} \bm{\psi} + \mathrm{h.c.} \big)\hspace{0.2mm}, %= -\, \frac{1}{2} \psi_{i I \alpha} \mathcal{M}_{ij} \delta_{IJ} \epsilon^{\alpha\beta} \psi_{j J \beta} + \mathrm{h.c.}\,.
\end{equation}
where $\bar{\sigma}^\mu \equiv (1, -\hspace{0.2mm}\sigma^i)$ and $\epsilon \equiv i\sigma^2$, and the covariant derivative $D_\mu \equiv \partial_\mu + i g A_\mu$ includes all the gauge potentials acting on the corresponding field. We ignore the topological $\theta$-term of the dark gauge group, whose effects in QCD-like dark sectors have been considered recently in Refs.~\cite{Garcia-Cely:2024ivo,Garcia-Cely:2025flv}.

To ensure that some of the pNGBs are stable, we assume that the $SO(2) \simeq U(1)$ unbroken flavor symmetry generated by $T_2 = \mathrm{diag}\,(\sigma_2, 0)/2$ is exact.\footnote{In an underlying UV completion, $U(1)_{T_2}$ could be a gauge symmetry, with the associated dark photon $\hat{\gamma}$. The symmetry could be unbroken and very weakly coupled; in this case the anomalous $\he \to \hat{\gamma} \hat{\gamma}$ decay would be present, since $\mathrm{Tr}\,[X_0 (T_2)^2] \neq 0$. Alternatively, it could be broken in such a way as to preserve a discrete symmetry, which suffices to protect the stability of the charged states.} This implies that the quark mass matrix takes the form
\begin{equation}
\mathcal{M} = \mathrm{diag}\;(m_1, m_1, m_3) \,.
\end{equation}
Under a $U(1)_{T_2}$ transformation, the linear combinations $\psi_{\pm 1/2} \equiv (\psi_1 \mp i \psi_2)/\sqrt{2}$ have charges $\pm 1/2$, while $\psi_0 \equiv \psi_3$ has charge $0$. 

The Goldstone matrix is defined as $\widehat{\Pi} = \hpi_i X_i$, while $\Sigma = e^{i\widehat{\Pi}/f_{\hpi}} e^{i\widehat{\Pi}^T/f_{\hpi}} = e^{i 2 \widehat{\Pi}/f_{\hpi}}$ transforms as $\Sigma \to U \Sigma U^T$. The normalization corresponds to the $f_{\pi} \approx 92$~MeV convention in the SM. It is convenient to classify the pNGBs as the charge eigenstates under $U(1)_{T_2}$, namely 
\begin{equation}
\hpi_{\pm} = \frac{1}{\sqrt{2}}(\hpi_3 \mp i \hpi_1),\qquad \hK_{\pm 1/2} = \frac{1}{\sqrt{2}}(\hpi_4 \mp i \hpi_6),\qquad \he_0 = \hpi_8. 
\end{equation}
Making use of the spurionic transformation property $\mathcal{M} \to U^\ast \mathcal{M} U^\dagger$, we write the $\mathcal{O}(p^2)$ terms of the chiral perturbation theory (ChPT) Lagrangian as
\begin{equation} \label{eq:p2}
\mathcal{L}_{\hpi, \,p^2} = \frac{f_{\hpi}^2}{4} \mathrm{Tr} \,[\partial_\mu \Sigma^\dagger \partial^\mu \Sigma ] + \frac{B f_{\hpi}^2}{2} \mathrm{Tr}\,[\mathcal{M}\Sigma + \Sigma^\dagger \mathcal{M}^\dagger ]\,,
\end{equation}
which gives the pNGB masses
\begin{equation}\label{eq:pion_masses_LO}
m^2_{\hat{\pi}} = 2 B m_1\,,\qquad m^2_{\hK} = B  (m_1 + m_3)\,, \qquad m^2_{\tilde{\eta}} = \frac{2B}{3}  (m_1 + 2 m_3 )\,.
\end{equation}
Assuming $m_3 > m_1$ realizes the mass hierarchies $m_{\hpi} < m_{\hK} < m_{\he}$ and ensures that both $\hat{\pi}_{\pm}$ and $\hat{K}_{\pm 1/2}$ are stable due to $U(1)_{T_2}$ charge conservation. Across most of the parameter space, the lighter $\hpi_{\pm}$ will be the dominant component of DM. By contrast, $\hat{\eta}_0$ is not protected by any symmetry and will decay in general. Its quantum numbers are $J^{PC} = 0^{-+}$.

The leading-order (LO) masses in Eq.~\eqref{eq:pion_masses_LO} satisfy the relation $3m_{\he}^2 + m_{\hpi}^2 = 4m_{\hK}^2$. The mass hierarchy $2\hspace{0.2mm} m_{\hK} > m_{\he} + m_{\hpi}$ is also satisfied, hence the scattering process $\hK_{\pm 1/2} \hK_{\pm 1/2} \to \hpi_{\pm} \he_0$ is always kinematically open, with important phenomenological consequences that we will discuss carefully. We expect this mass hierarchy to hold even beyond LO, where the most important correction will only reduce the $m_{\he}$: this is the mixing between $\he_0$ and the heavier $\he'$, the $SU(N_f)\,$-$\,$singlet meson associated with the $U(1)$ acting as $\bm{\psi}\to e^{i\alpha}\bm{\psi}$, which is anomalous with respect to $SO(N_d)$.

The chiral Lagrangian predicts the four-point interactions among the pNGBs in terms of $f_{\hpi}$ and the masses in Eq.~\eqref{eq:pion_masses_LO}. The $\hpi^4, \hK^4, \hK^2 \hpi^2 , \he^2 \hK^2$ and $\hK^2 \he \hpi$ couplings are mediated by both derivative and mass terms, while $\he^4$ and $\he^2 \hpi^2$ are only mediated by mass terms (henceforth the $U(1)_{T_2}$ charge subscripts are understood, unless confusion can arise). These interactions are responsible for $2\to 2$ scattering processes. Among these processes, those that change dark meson species will play a crucial role in determining the DM relic density. All of them proceed in the $s$-wave and, aside from possible kinematic suppressions due to small mass splittings, scale as $\langle \sigma v \rangle \propto m^2/f_{\hpi}^4$ where $m$ generically denotes the pNGB masses. In detail, the following pattern of thermally-averaged cross sections is predicted
\begin{align}\label{eq:2to2_hierarchies}
\langle \sigma_{\he\he\to\hK\hK} v\rangle  > \langle \sigma_{\hK\hK\to\hpi\he} v \rangle > \langle  \sigma_{\hK\hK\to \hpi\hpi} v \rangle > \langle \sigma_{\hK\he\to\hK\hpi} v \rangle \gg \langle \sigma_{\he\he\to\hpi\hpi} v \rangle\,,
\end{align}
with complete expressions provided in Appendix~\ref{app:Boltzmann}.

The $SU(3)/SO(3)$ coset also contains a Wess-Zumino-Witten~(WZW) term~\cite{Hochberg:2014kqa}, which can be written as
\begin{align}\label{eq:WZW}
\mathcal{L}_{\hpi,\,\rm WZW} = \frac{2N_d}{15\pi^2 f_{\hpi}^5} \frac{5\sqrt{3}}{8} \,\epsilon^{\mu\nu\rho\sigma} \he_0 \partial_\mu \hK_{+1/2} \partial_\nu \hK_{-1/2} \partial_\rho \hpi_+ \partial_\sigma \hpi_- \,.
\end{align}
%where $\mathcal{T}_{86431} = 60\, \mathrm{Tr}\,[X_8 X_6 X_4 X_3 X_1] = 5\sqrt{3}/8$. 
The WZW term mediates a variety of $3\to 2$ processes, such as \eg $\hpi_+ \hpi_- \hK_{\pm 1/2} \to \he_0 \hK_{\pm 1/2}$, which deplete the number density of $\hat{\pi}$. Such interactions determine the DM relic density in SIMP scenarios~\cite{Hochberg:2014dra,Hochberg:2014kqa,Hochberg:2015vrg} (see also Refs.~\cite{Carlson:1992fn,Machacek:1994vg,deLaix:1995vi} for pioneering studies). However, given the steep dependence of the thermally-averaged cross section on $m/f_{\hpi}$, $\langle \sigma v^2 \rangle \propto N_d^2 m^3 T^2/f_{\hpi}^{10}$, $3\to 2$ processes play an important role in setting the pNGB abundances only for large values of $m/f_{\hpi}$, not far from the non-perturbative limit of $4\pi$. In this work we focus on the perturbative regime $m/f_{\hpi} \lesssim \mathcal{O}(1)$, hence WZW interactions will not play any significant role (though we include them for completeness, both in our analytical and numerical results). 

Taking $m/f_{\hpi} \lesssim \mathcal{O}(1)$ also ensures that the dominant component of DM, $\hpi$, self-scatters with a cross section consistent with bounds from DM halo shapes and merging galaxy clusters. For instance, observations of the Bullet cluster require $\sigma/m < \mathcal{O}(1)\;\mathrm{cm}^2/\mathrm{g}$~\cite{Tulin:2017ara}. The LO chiral Lagrangian gives, after averaging over same-sign and opposite-sign scatterings, the velocity-independent cross section
\begin{equation}
\frac{\sigma_{\hpi \hpi \to \hpi \hpi}}{m_{\hpi}} \simeq  \frac{3m_{\hpi}}{64\pi f_{\hpi}^4} \approx 3.3 \times 10^{-6}\, \frac{\mathrm{cm}^2}{\mathrm{g}}\,\bigg( \frac{\mathrm{GeV}}{m_{\hpi}}  \bigg)^3 \bigg( \frac{m_{\hpi}/f_{\hpi}}{1} \bigg)^4   \,,
\end{equation}
which is safely allowed in the whole mass range considered in this work, $m_{\hpi}\gtrsim 100\;\mathrm{MeV}$.

Having characterized the light dark meson states,\footnote{The dark baryons~\cite{Antipin:2015xia} are built as antisymmetric combinations of $N_d$ dark quarks. For even $N_d$, the lightest baryon is plausibly a spin-0 state singlet under $SO(N_f)$, whereas for odd $N_d$, it is a spin-$1/2$ state $\bm{\Psi}$ transforming in the fundamental representation of $SO(N_f)$. In turn, the latter splits into $\Psi_{\pm 1/2}$ and $\Psi_0$, where the subscript denotes $U(1)_{T_2}$ charges. There is no conserved baryon number, yet the lightest baryon is stable due to a $Z_2 = O(N_d)/SO(N_d)$ symmetry that is accidentally preserved by the $SO(N_d)$ gauge theory. The dark baryons annihilate very efficiently in the early Universe, leading to a negligible present-day abundance.} next we introduce interactions with the SM.

\subsection{Portal to the Standard Model}\label{sec:portal}
We assume the existence of a portal interaction which allows $\he$ to decay to SM particles, keeping the dark and visible sectors in thermal equilibrium in the early Universe. The $\he$ decays deplete the overall number density of dark pNGBs and make it possible for the lightest stable species $\hat{\pi}$ to acquire a relic abundance that matches the observed amount of DM (with a subleading contribution from the heavier $\hK$). This mechanism for thermal DM production can be realized through different portals, because the relic densities of the dark pNGBs are sensitive only to the total width $\Gamma_{\he}$, but not to the details of the coupling between the dark and visible sectors. 

The nature of the portal is, however, crucial to determine the signals in accelerator experiments, as well as direct and indirect DM searches, that could be exploited to discover the dark sector. The pseudoscalar $\he$ can be viewed as a light composite axion-like particle (ALP). Hence, its leading low-energy interactions with the light SM fields arise at dimension $5$ and take the form $\partial_\mu \he_0 \bar{f}\gamma^\mu \gamma_5 f$, where $f$ denotes the SM fermions, or $\he_0 V_{\mu\nu} \widetilde{V}^{\mu\nu}$ with $V = G, F$ (barring $CP$-violating effects, which we do not consider in this work). In Sec.~\ref{sec:UV} we present several UV completions that give rise to couplings of $\he$ to SM fermions, and discuss the most promising experimental strategies to probe them. 

One important feature that all these UV completions have in common, is that they preserve the discrete charge conjugation symmetry $\mathcal{C} \in O(N_f)$ associated to $U(1)_{T_2}$, which acts as $\psi_i \stackrel{\mathcal{C}}{\to} -(-1)^i\, \psi_i$ on the dark quarks ($i=$ 1, 2, 3) and as $\hpi_{\pm} \to \hpi_{\mp}$, $\hK_{\pm 1/2} \to \hK_{\mp 1/2}$ and $\he_0 \to \he_0$ on the dark mesons. As a consequence, interactions of the form $i(\hpi_+ \partial_\mu \hpi_- - \hpi_- \partial_\mu \hpi_+ ) J_{\rm SM}^\mu$ are forbidden, since the DM vector current is odd under $\mathcal{C}$ while $J_{\rm SM}^\mu$, denoting a generic vector current made of SM fermions, does not transform under $\mathcal{C}$. Some UV completions do generate $\mathcal{C}$-invariant {\it scalar} current interactions, of the form $\hpi_+ \hpi_- J_{\rm SM}$, but these are strongly suppressed by the SM fermion masses. As a result, the scenario we consider avoids constraints from direct and indirect DM searches in most of its parameter space. This is especially important for direct detection, where vector interactions involving SM quarks would easily conflict with data; the discrete $\mathcal{C}$ symmetry robustly avoids this.\footnote{If $U(1)_{T_2}$ is gauged by a dark photon $\hat{A}_\mu$, $\mathcal{C}$-invariance also forbids a kinetic mixing operator with the SM hypercharge, $\hat{F}_{\mu\nu} B^{\mu\nu}$, since $\hat{A}_\mu \stackrel{\mathcal{C}}{\to} -\, \hat{A}_\mu$ while $B_\mu$ does not transform under $\mathcal{C}$.} For these reasons, searches at accelerators emerge as the most promising avenue to discover this class of thermal GeV-scale dark sectors. As a matter of fact, in large regions of parameter space, accelerators offer the {\it only} viable approach to discovery. We show concrete examples of this in Sec.~\ref{sec:UV}. 

Before turning to the thermal evolution, we elaborate on our choice to focus in this work on the mass range $0.1 \lesssim m_{\hpi}/\mathrm{GeV} \lesssim 10$, where high-multiplicity dark shower events~\cite{Albouy:2022cin} are generically expected to arise at colliders. For dark pNGBs lighter than $2m_\mu \approx 0.2\;\mathrm{GeV}$, explicit UV completions show that the expected $\he$ lifetime would be too long to effectively deplete the dark sector number density and achieve the observed DM relic abundance. Conversely, for masses larger than 10 GeV the observed DM abundance can still be satisfied, with smaller mass splittings. However, the expected multiplicity in collider events is strongly suppressed and final states containing only a few dark pNGBs are expected, instead of the dark shower topologies we focus on here. For previous studies of the heavier mass range, see Refs.~\cite{Beauchesne:2018myj,Beauchesne:2019ato}.

\section{Dark Sector Thermal Evolution and Dark Matter Relic Density}
\label{sec:thermal}
The thermal evolution of the dark sector is governed by two classes of processes: $2\to 2$ scatterings among the pNGBs $\hat{\pi}, \hat{K}$ and $\hat{\eta}$, and decays and inverse decays of $\he$ to SM particles. Initially, the $2\to 2$ scatterings, which have rather large cross sections, keep all pNGBs in chemical and kinetic equilibrium. Furthermore, efficient (inverse) decays ensure that the dark sector and the SM are kept in chemical equilibrium and share a single temperature $T$. As we discuss below, we focus on parameter regions where the $\he$ decay width is large enough to keep the temperatures of the two sectors equal at least until the process that determines the relic density of $\hat{\pi}$, which is the dominant DM component, decouples. Therefore, a Boltzmann description in terms of a single temperature $T$, or $x \equiv m_{\hat{\pi}}/T$, suffices to determine the DM relic density to good accuracy.\footnote{Notice that, due to the dark charge conjugation $\mathcal{C}$ discussed in Sec.~\ref{sec:portal}, scattering of dark pNGBs with SM particles is generically highly suppressed and not efficient in maintaining kinetic equilibrium between the dark and visible sectors.} Since we take $m/f_{\hpi} \lesssim \mathcal{O}(1)$, the $3\to 2$ processes mediated by the WZW term in Eq.~\eqref{eq:WZW} freeze out early, at $x\lesssim 10$, and play a negligible role. 

Four parameters control the thermal evolution: the total decay width of $\he$, $\Gamma_{\he}$; the DM mass, $m_{\hpi}$; the mass splitting $\Delta_{\he} \equiv m_{\he}/m_{\hpi} - 1$, which also fixes the other splitting according to the LO ChPT prediction,
\begin{equation} \label{eq:DeltaK_LO}
\Delta_{\hK} \equiv \frac{m_{\hK}}{m_{\hpi}} - 1 = \bigg[1+ \frac{3}{4} \Delta_{\he} (2 + \Delta_{\he}) \bigg]^{1/2} - 1\,;
\end{equation}
and the ratio $m_{\hpi}/f_{\hpi}$, which determines the strength of the interactions among the pNGBs.\footnote{The number of dark colors, $N_d$, enters the coefficient of the WZW term which governs $3\to 2$ processes. As already mentioned, however, the latter freeze out early and play an insignificant role in the thermal evolution.} Starting from the ChPT Lagrangian discussed in Sec.~\ref{sec:model}, we derive Boltzmann equations that describe the evolution of the comoving number densities of the dark sector species, $Y_X \equiv n_X / s$ with $X = \hpi, \hK, \he$, where $s$ is the total entropy density and we define $Y_{\hpi} = Y_{\hpi_+} = Y_{\hpi_-}$ and $Y_{\hK} = Y_{\hK_{+1/2}} = Y_{\hK_{-1/2}}$. The complete Boltzmann equations are reported in Appendix~\ref{app:Boltzmann}. In this section, we analyze their central features, emphasizing analytical insight wherever possible.

After $3\to 2$ processes have frozen out, the total (comoving) dark sector number density only changes due to decays of $\he$,
\begin{equation}
\label{eq:Gammaeff2}
\frac{d Y_{\rm tot} }{dx}  \simeq  -\, \frac{\Gamma_{\he}}{\tilde{H}(x) x}  \big(Y_{\he} - Y_{\he}^{\rm eq} \big)~, \qquad Y_{\rm tot} \equiv 2 Y_{\hpi} + 2 Y_{\hK} + Y_{\he}\,, 
\end{equation}
where $\tilde{H}(x) = H(x) / \big( 1 - \tfrac{1}{3}\frac{d \log g_{\ast s}}{d\log x}\big)$, with $g_{\ast s}$ the effective number of degrees of freedom for entropy density, to which the dark sector contributes negligibly. Chemical equilibrium with the SM is kept as long as the $\he$ decay rate is big enough to reduce the number of $\hpi$ and $\hK$ efficiently. In the instantaneous decoupling approximation, this ceases to hold when
\begin{equation}\label{eq:Gammaeff}
\Gamma_{\rm eff}(x)\equiv \frac{Y_{\he}^{\rm }(x)}{Y_{\rm tot}(x)}\Gamma_{\he}  \simeq x\hspace{0.15mm} H(x)\,,
\end{equation}
where the effective width $\Gamma_{\rm eff} \simeq Y_{\he}\hspace{0.2mm} \Gamma_{\he}/(2 Y_{\hpi})$ accounts for the fact that $\hpi$ (and $\hK$) is much more abundant than $\he$. The DM relic density is determined either by the decoupling of the effective decay rate in Eq.~\eqref{eq:Gammaeff}, or by the freeze-out of $\hpi\to \he$ conversions via $2\to 2$ scatterings, whichever happens {\it earlier}. We term the latter situation ``large-$\Gamma_{\he}$ regime'' and the former ``small-$\Gamma_{\he}$ regime.'' 

As a result, in all the parameter space where the observed DM abundance is produced by the freeze-out of $\hpi$, the ratio $\Gamma_{\rm eff}/(xH)$ becomes smaller than $1$ at $x_{\Gamma} \sim 20$ (in the small-$\Gamma_{\he}$ regime) or later (in the large-$\Gamma_{\he}$ regime). However, this does not automatically ensure that the dark and SM sectors shared the same temperature at all earlier times, because $\Gamma_{\rm eff}/(xH)$ is not monotonically decreasing with time. While it does decrease rapidly for $x \gg \Delta_{\he}^{-1}$ due to the Boltzmann suppression of $\Gamma_{\rm eff}$, for $x \lesssim \Delta_{\he}^{-1}$ it actually increases with $x$, as $H\propto x^{-2}$ while the Boltzmann suppression has not kicked in yet. To be conservative, we therefore impose an additional condition, restricting our focus on regions of parameter space that satisfy
\begin{equation}\label{eq:kin_eq_lost_early}
    \frac{\Gamma_{\rm eff}(x)}{x H(x)}\bigg|_{x\, =\, 1} > 1 \qquad \to \qquad \Gamma_{\he} \gtrsim 5\, \frac{\pi  g_\ast (T = m_{\hpi})^{1/2} m_{\hpi}^2}{3\sqrt{10}\, M_{\rm Pl}}\,.
\end{equation}
To obtain the second form of the inequality in Eq.~\eqref{eq:kin_eq_lost_early} we have approximated $Y_{\he}^{\rm eq}/Y_{\rm tot}^{\rm eq} \simeq 1/5$ at $x = 1$, when it is still acceptable to take the ultra-relativistic limit of $Y^{\rm eq}$. Together with the plausible assumption that at even higher temperatures ($x < 1$) the kinetic equilibrium was kept by scatterings of dark quarks (or, below the critical temperature of the dark confinement transition, non-pNGB dark hadrons) with light SM particles, the condition in Eq.~\eqref{eq:kin_eq_lost_early} ensures that a description in terms of a single temperature $T$ is appropriate until DM freeze-out. 

We now turn to present the salient features of the large-$\Gamma_{\he}$ and small-$\Gamma_{\he}$ regimes. Related dynamics was discussed in Refs.~\cite{Beauchesne:2018myj,Beauchesne:2019ato,Bernreuther:2019pfb,Li:2019ulz,Frumkin:2021zng,Frumkin:2025dxq}.

\subsection{The large-$\Gamma_{\he}$ and small-$\Gamma_{\he}$ regimes}\label{sec:two_regimes}

\begin{figure}[t]
    \centering
    \includegraphics[width=0.49\linewidth]{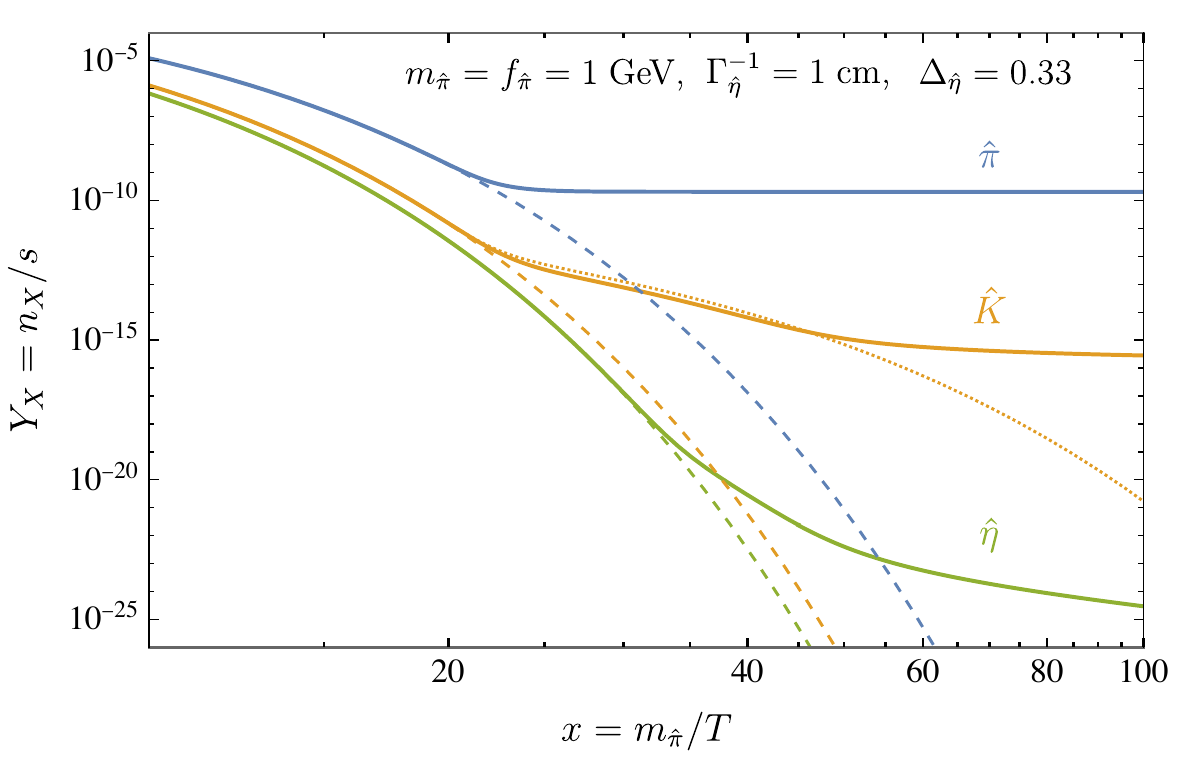}\hspace{1mm}
    \includegraphics[width=0.49\linewidth]{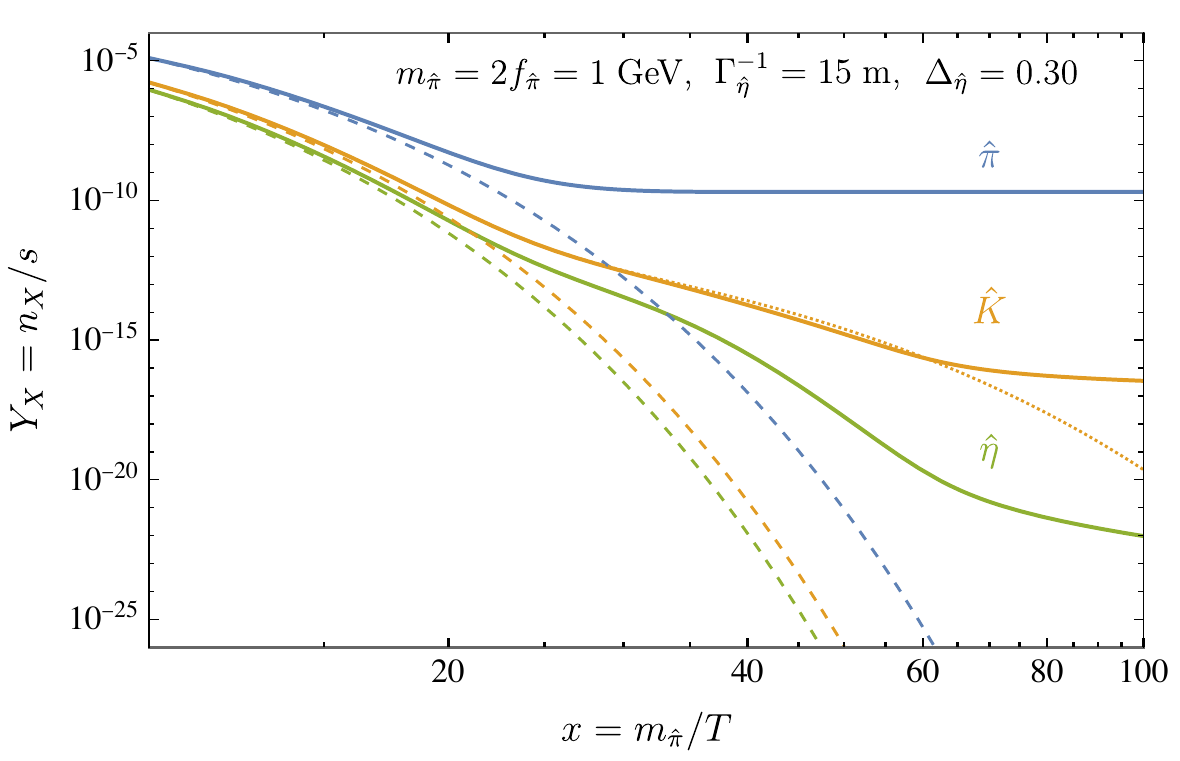}
    \caption{Evolution of the comoving number densities of the dark meson species. The DM mass is fixed to $m_{\hpi} =1\;\mathrm{GeV}$ in both panels. Solid curves show numerical solutions to the Boltzmann equations, while dashed curves correspond to the equilibrium abundances $Y_{X}^{\rm eq}$. The dotted orange curve shows $Y_{\hpi} \big(Y_{\hK}^{\rm eq}/Y_{\hpi}^{\rm eq}\big)$, see the text for more details. {\it Left:} The large-$\Gamma_{\he}$ regime, where the DM relic density is determined by the freeze-out of $\hpi \to \he$ conversions via $2\to 2$ scattering processes. We choose $\Gamma_{\he}^{-1} = 1\;\mathrm{cm}$ and set the remaining parameters to $f_{\hpi} = 1\;\mathrm{GeV}$ and $\Delta_{\he} = 0.33$ to reproduce the observed DM abundance. {\it Right:} The small-$\Gamma_{\he}$ regime, where the DM relic density is determined by the loss of chemical equilibrium with the SM via $\he$ decays. We choose $\Gamma_{\he}^{-1} = 15\;\mathrm{m}$ and set $f_{\hpi} = 0.5\;\mathrm{GeV}$ and $\Delta_{\he} = 0.30$ to match the observed DM abundance.}
    \label{fig:main_scenarios}
\end{figure}

An example of the large-$\Gamma_{\he}$ regime is shown in the left panel of Fig.~\ref{fig:main_scenarios}. In this case, $\Gamma_{\rm eff}$ is large enough to keep all dark sector species in equilibrium until $\hpi$ freezes out. Among the flavor-changing up-scattering processes that convert $\hpi$ to $\he$, $\hpi\hpi \to \he\he$ freezes out first, due to the smaller cross section 
%(see Eq.~\eqref{eq:2to2_hierarchies}) 
and stronger Boltzmann suppression. The other processes freeze out at similar times, with $\hK \hpi \to  \hK \he$ decoupling slightly earlier. 
The key process that governs the overall freeze-out is $\hpi\hpi\to \hK\hK$ (combined with $\hK\hK\to\hpi\he$, which is always kinematically open). 
In this example, the final $\hpi$ density corresponds to a freeze-out temperature $x_f\approx 22.4$. An analytic estimate of the freeze-out temperature from the detailed balance equation, 
\begin{equation}\label{eq:FO_large_width}
Y_{\hpi}^{\rm eq} (x)\, e^{- 2 \Delta_{\hK} x}\, \langle \sigma v \rangle_{\hK\hK\, \to\, \hpi\hpi} \simeq \frac{H(x)}{s(x)}\,,
\end{equation}
gives $x_f \sim 24$. This is a decent but not very accurate approximation, as typical for the freeze-out of forbidden annihilation processes. 

After $\hpi$ freezes out, the processes $\hK\hK\leftrightarrow \hpi\hpi$ and $\hK\hK\leftrightarrow \hpi\he$ still keep $\hK$ in chemical equilibrium with $\hat{\pi}$, though with non-zero chemical potentials $\mu_{\hK} \simeq \mu_{\hpi} \gg \mu_{\he}$: the comoving number densities satisfy $Y_{\hK} \simeq Y_{\hpi} \big(Y_{\hK}^{\rm eq}/Y_{\hpi}^{\rm eq}\big)$ where $Y_{\hpi} = Y_{\rm DM}/2$, until these processes also freeze out around $x_{f}^{\hK} \approx 50$. 
The $\hpi$ abundance is negligibly affected by the injections produced by $\hK\hK\to \hpi\hpi$ and $\hK\hK\to \hpi\he$, since $Y_{\hK} \ll Y_{\hpi}$. On the other hand, after $x_f$ the $\he$ maintains an equilibrium distribution for a while, before departing from equilibrium when the decay rate of $\he$ to SM particles becomes smaller than the rate of injections from $\hK \hK \to \hpi \he$.  Note that the evolution of the $\hpi$, $\hK$, and $\he$ densities in Fig.~\ref{fig:main_scenarios} assumes that the dark and visible sectors retain the same temperature $T$ throughout. However, when the equality in Eq.~\eqref{eq:Gammaeff} is reached the two sectors chemically decouple (this happens at $x_\Gamma \approx 27$, for the example shown in the left panel of Fig.~\ref{fig:main_scenarios}), and we expect kinetic decoupling to happen shortly after. Then, the subsequent dynamics of $Y_{\hK}$ and $Y_{\he}$ will be altered, since the temperature of the dark sector drops faster, $\hat{T} \propto a^{-2}$. Although this may have some impact on the final abundance of $\hK$ (which is anyway small in most of our parameter space), it does not affect the abundance of the dominant DM component $\hpi$, which remains a robust outcome of our calculations.

The right panel of Fig.~\ref{fig:main_scenarios} shows an example of the small-$\Gamma_{\he}$ regime. In this case, chemical equilibrium with the SM is lost when $\hpi$ up-scattering processes (\eg $\hpi\hpi \to \hK\hK$) are still in equilibrium. Once the equality in Eq.~\eqref{eq:Gammaeff} is reached (at $x_\Gamma \sim 20$, for the chosen example), $Y_{\he}$ deviates from the equilibrium value, which almost simultaneously drives $Y_{\hpi}$ and $Y_{\hK}$ away from equilibrium, as well. This effectively determines the freeze-out of the total dark meson abundance, which is dominated by the $\hpi$  component. The decoupling of the decay process is rather slow, hence the accuracy of the relic abundance estimate obtained from the instantaneous freeze-out approximation, $Y_{\hpi}^{\rm eq}(x_\Gamma)$, is limited. For $x > x_\Gamma$, all dark sector species evolve with chemical potentials $\mu_{\hK} \simeq \mu_{\hpi} \gtrsim \mu_{\he}$ until the scatterings $\hK\hK \to \hpi\hpi$ and $\hK\hK \to \hpi\he$ freeze out. As noted earlier, after $x_\Gamma$ the dark sector is expected to develop a different temperature $\hat{T}\neq T$, but the modified dynamics of the two heavier species will have a negligible impact on the prediction of the $\hpi$ relic abundance.

To conclude, we briefly comment on more general mass spectra where $\Delta_{\he}$ and $\Delta_{\hK}$ deviate from the LO ChPT prediction in Eq.~\eqref{eq:DeltaK_LO}. In the large-$\Gamma_{\he}$ regime, DM freeze-out is determined by $\hpi \hpi \to \hK\hK$ scattering according to Eq.~\eqref{eq:FO_large_width}, hence the DM relic abundance is sensitive to $\Delta_{\hK}$ instead of $\Delta_{\he}$ as long as $2m_{\hK} > m_{\hpi} + m_{\he}$ remains true. Conversely, in the small-$\Gamma_{\he}$ regime it is the decay of $\he$ to SM particles that determines the freeze-out, which is set by $\Delta_{\he}$ via Eq.~\eqref{eq:Gammaeff}.

\subsection{Viable parameter space}

\begin{figure}[t]
\centering
\includegraphics[width=0.48\textwidth]{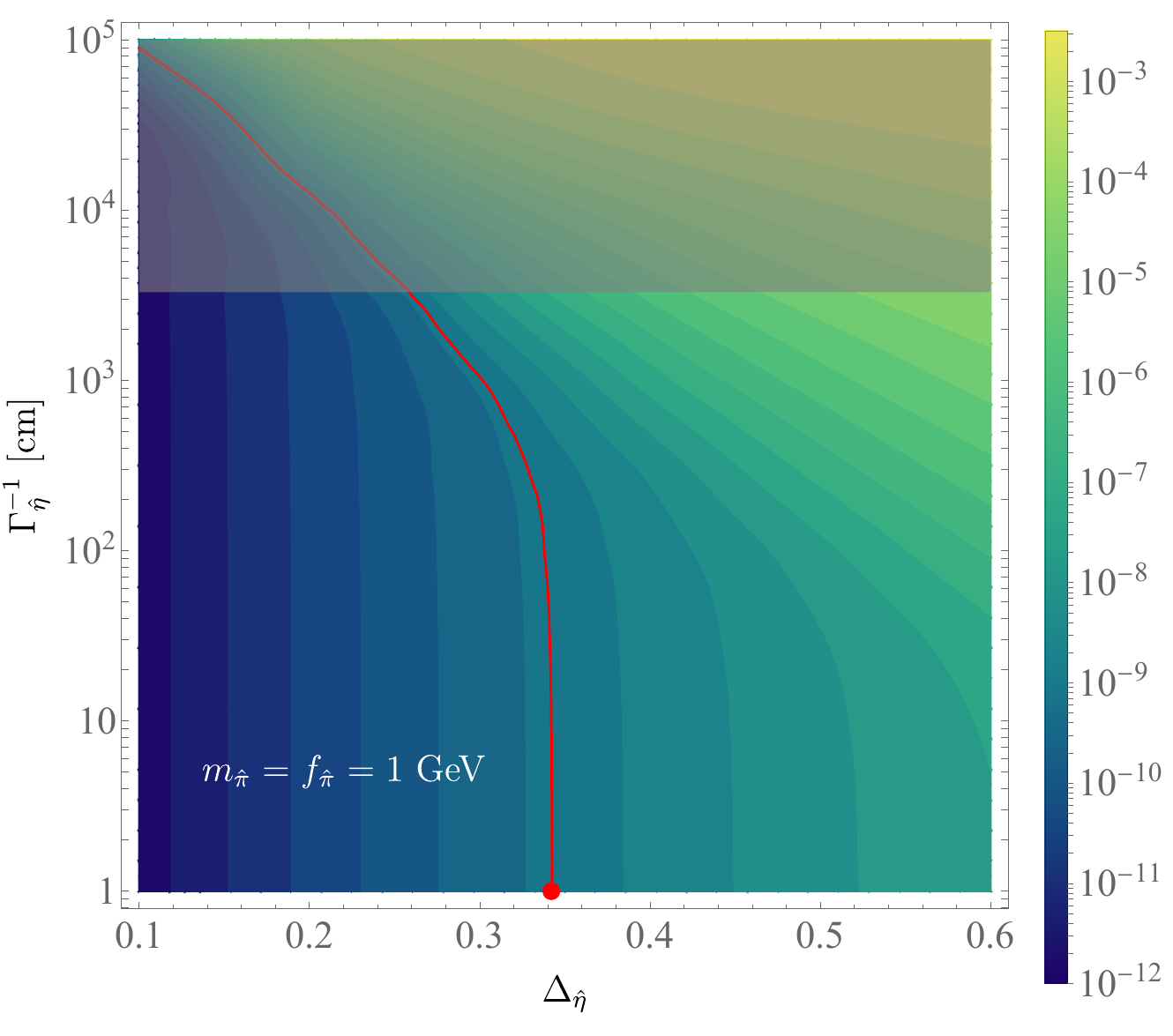}\hspace{2mm}
\includegraphics[width=0.48\textwidth]{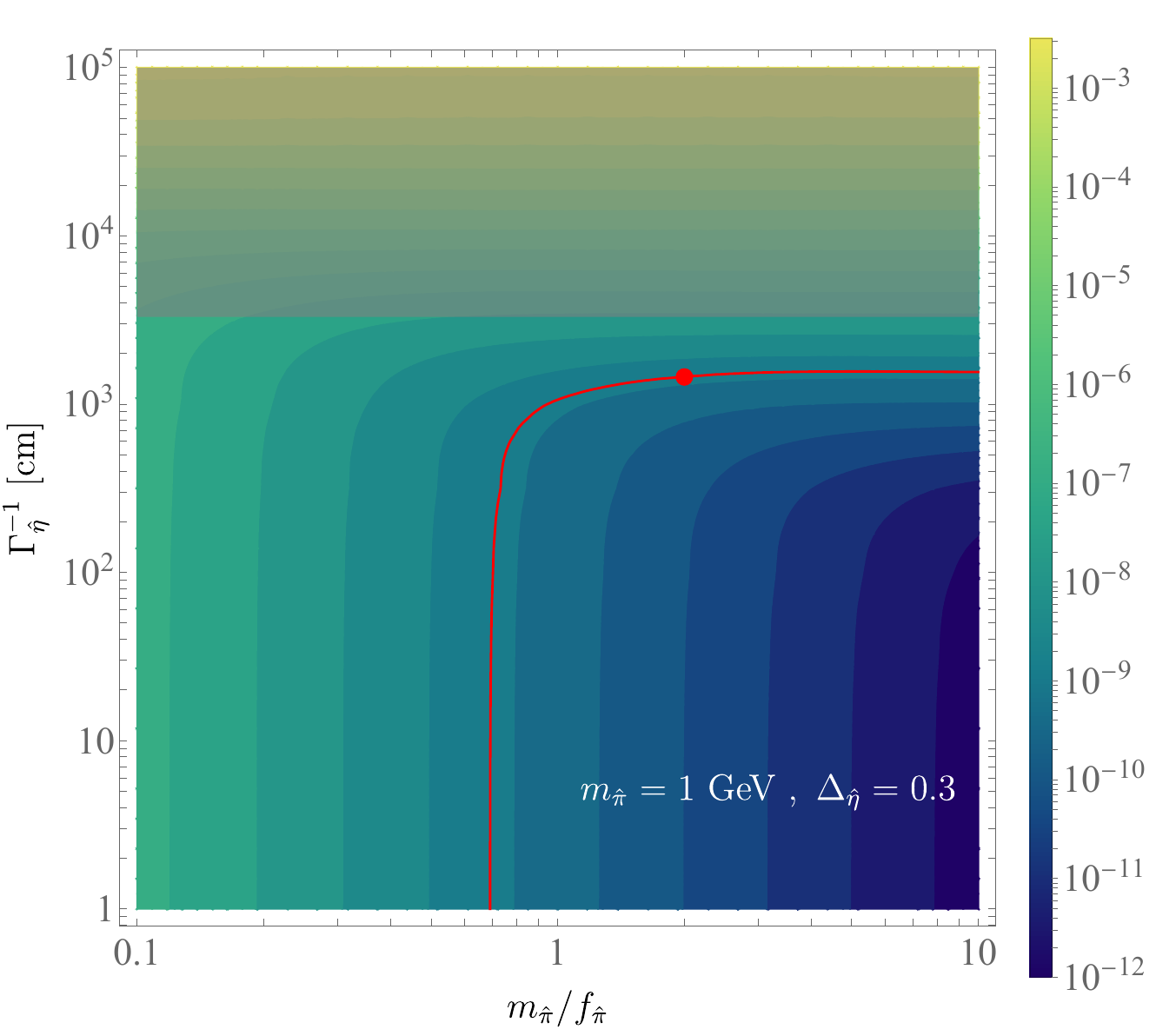}
\caption{{\it Left:} Isocontours of $2Y_{\hpi}^0$ in the $(\Delta_{\he}, \Gamma_{\he}^{-1})$ plane, for $m_{\hpi} = f_{\hpi} = 1$~GeV. The red curve corresponds to the observed DM abundance. {\it Right:} Similar isocontours, but in the $(m_{\hpi}/f_{\hpi}, \Gamma_{\he}^{-1})$ plane, fixing $m_{\hpi} = 1$~GeV and $\Delta_{\he}=0.3$. In both panels, a gray shading covers the region where $\Gamma_{\he}$ is too small to guarantee that the dark and SM sectors share a common temperature at all times before DM freeze-out, see Eq.~\eqref{eq:kin_eq_lost_early}. The red dots correspond to the examples exhibited in Fig.~\ref{fig:main_scenarios}.}
\label{fig:param1}
\end{figure}

The dependence of the DM yield on $\Gamma_{\he}$ and the other model parameters is shown in Fig.~\ref{fig:param1}. Red contours indicate where the $\hpi_{\pm}$ abundance matches $\Omega_{\rm DM} h^2 = 0.120$~\cite{Planck:2018vyg}; the contribution from $\hK_{\pm 1/2}$ is always negligible in the shown parameter space. Regions where $2Y_{\hpi}^0$ is larger than the observed DM abundance are ruled out, unless a modified cosmological history is invoked. By contrast, regions where $\hat{\pi}$ is under-abundant can be viable even for standard cosmology, if some other candidate supplies the missing DM energy density.

\begin{figure}[t]
\centering
\includegraphics[width=0.45\textwidth]{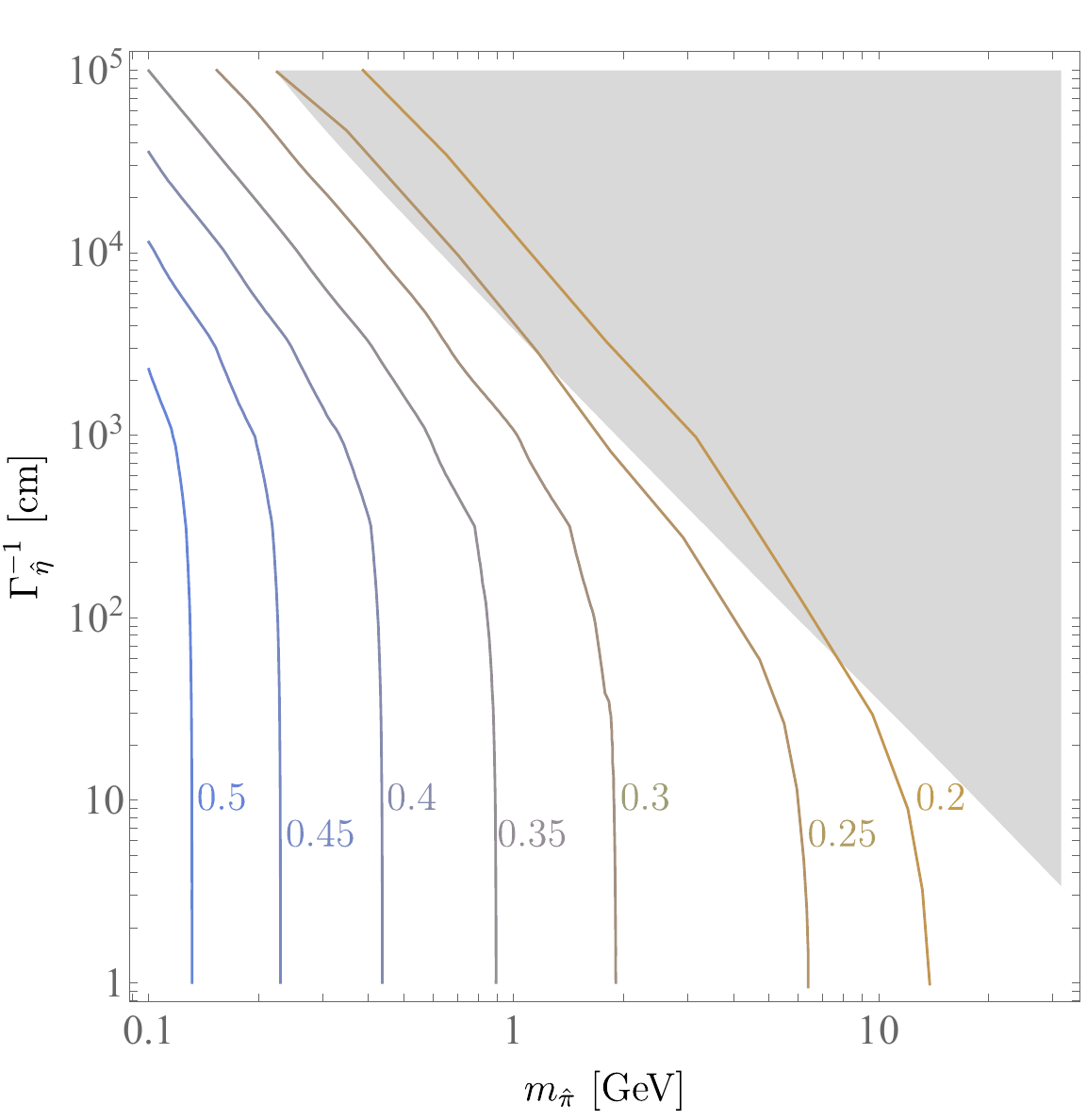}\hspace{3mm}
\includegraphics[width=0.45\textwidth]{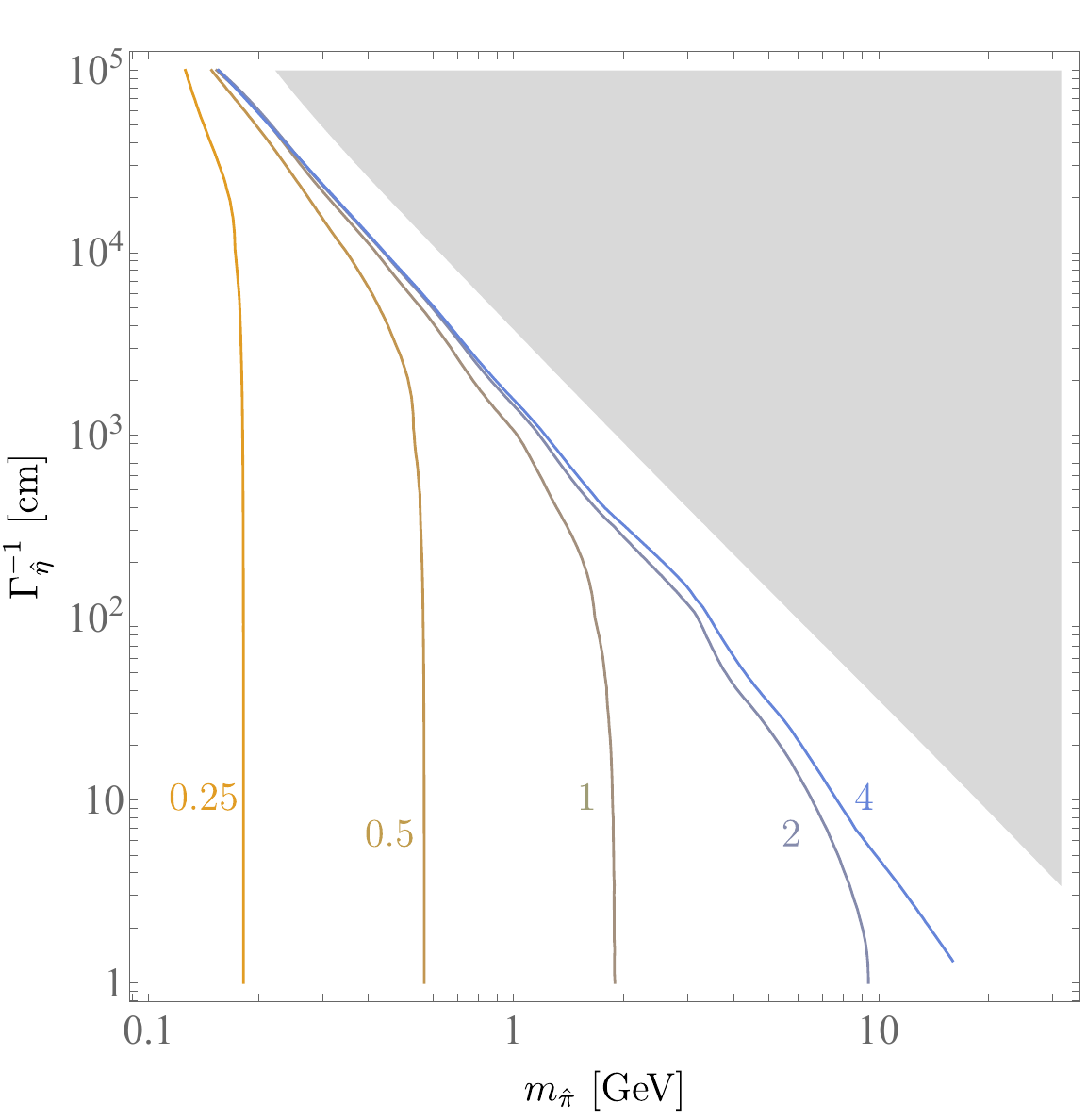}
\caption{{\it Left:} Contours where the $\hpi_{\pm}$ relic density matches the observed DM density, for $m_{\hpi}/f_{\hpi} = 1$ and a few representative values of $\Delta_{\he}$. {\it Right:} Similar, but fixing $\Delta_{\he}=0.3$ and taking a few values of $m_{\hpi}/f_{\hpi}$. 
In both panels, we shaded in gray the region where $\Gamma_{\he}$ is too small to guarantee that the dark and SM sectors share a common temperature at all times before DM freeze-out, see Eq.~\eqref{eq:kin_eq_lost_early}.
}
\label{fig:param3}
\end{figure}

For each contour in Fig.~\ref{fig:param1}, the critical point marking the transition between the large-$\Gamma_{\he}$ and small-$\Gamma_{\he}$ regimes is clearly visible. In the left panel, we have fixed $m_{\hpi} = f_{\hpi} = 1\;\mathrm{GeV}$. For $\Gamma_{\he}$ larger than the critical value, the DM yield is determined by the freeze-out of $\hpi\to \he$ conversions and is therefore independent of the width, resulting in nearly-vertical contours. Conversely, for smaller $\Gamma_{\he}$ the DM yield increases with decreasing width, as the chemical decoupling between the dark and SM sectors happens earlier. Furthermore, increasing $\Delta_{\he}$ (and therefore $\Delta_{\hK}$ according to Eq.~\eqref{eq:DeltaK_LO}) always increases $Y_{\hpi}^0$. In the right panel, we have fixed $m_{\hpi} = 1\;\mathrm{GeV}$ and $\Delta_{\he} = 0.3$. For $\Gamma_{\he}$ larger than the critical value, $Y_{\hpi}^0$ is almost independent of the width, hence the contours are again vertical as already found in the left panel. On the other hand, for smaller $\Gamma_{\he}$ the DM yield is determined by the decoupling of the $\he$ decay according to Eq.~\eqref{eq:Gammaeff} and is therefore insensitive to $m_{\hpi}/f_{\hpi}$, which governs the conversion rates between the dark pNGBs. This results in nearly horizontal contours. 

In Fig.~\ref{fig:param3} we draw contours in the $(m_{\hpi}, \Gamma^{-1}_{\he})$ plane where the thermal relic abundance matches the observed value. In the left panel, we set $m_{\hpi}/f_{\hpi} = 1$ and consider several benchmark values for $\Delta_{\he}$. Larger $\Delta_{\he}$ leads to larger $Y^0_{\hpi}$ and therefore requires a smaller DM mass to match the observed DM energy density. The transition between the large- and small-$\Gamma_{\he}$ regimes is clearly visible on the contours, and depends mildly on $m_{\hpi}$. The region of parameter space where the $\he$ width is too small to satisfy the condition~\eqref{eq:kin_eq_lost_early} is shaded in gray. In the right panel, we fix $\Delta_{\he} = 0.3$ and take several benchmark values of $m_{\hpi}/f_{\hpi}$. Increasing $m_{\hpi}/f_{\hpi}$ implies stronger pNGB self-interactions. In the large-$\Gamma_{\he}$ regime this leads to a smaller $Y_{\hpi}^0$, hence the DM mass needs to be increased to reproduce the observed DM energy density. Conversely, in the small-$\Gamma_{\he}$ regime the value of $m_{\hpi}/f_{\hpi}$ has a minor impact on the DM relic density, as shown by the convergence of all the contours in the upper part of the plane.

\section{Irreducible Indirect Detection}\label{sec:ID}
Our framework predicts an irreducible indirect detection signal: since the $\hK$'s make up a sub-component of DM, they can annihilate in DM halos via the $\hK\hK \to \hpi \he$ process, followed by $\he$ decay to SM particles. Notice that the lifetime of $\he$ is very short relative to cosmological time scales, and in any case must satisfy $\Gamma_{\he}^{-1}\ll 1\,\mathrm{s}$ to be compatible with Big Bang Nucleosynthesis constraints, so we can view the decay as effectively instantaneous.

We can approximately repurpose the standard indirect detection constraints on DM annihilation to our $\hK \hK \to \hpi \he$ signal. For simplicity, we assume that $\he$ decays dominantly to a single two-body final state, $X_{\rm SM} \overline{X}_{\rm SM}$. We then require
\begin{equation}\label{eq:sigmaveff_1}
\xi^2 \langle \sigma v \rangle_{\hK \hK \to \hpi \he} \left( \frac{2 \rho_{\hK}^0}{\rho_{\rm DM}} \right)^2 \lesssim 2\, \langle \sigma v \rangle_{\rm obs} \Big(m_{\rm DM} = \xi m_{\hK} \Big)_{X_{\rm SM} \overline{X}_{\rm SM}}\,,
\end{equation}
where $\xi \equiv \big(4 m_{\hK}^2 + m_{\he}^2 -m_{\hpi}^2\big)\big/\big(8 m_{\hK}^2\big)$ is the fraction of the total energy $2m_{\hK}$ carried by the $\he$, assuming annihilation at rest ($\xi \approx 1/2$ for small mass splittings). In Eq.~\eqref{eq:sigmaveff_1}, the $\xi^2$ factor on the left-hand side corrects for the different number densities. $\langle \sigma v \rangle_{\rm obs}$ is the observed upper limit on $\langle \sigma v\rangle$ for present-day DM annihilation to $X_{\rm SM} \overline{X}_{\rm SM}$, obtained by assuming real DM with mass equal to $\xi m_{\hK}$, so that each $X_{\rm SM}$ is produced with the same energy as in our signal process (assuming annihilation at rest). The factor of $2$ on the right-hand side corrects for the fact that $\hK$ is not real, but complex. We can rewrite Eq.~\eqref{eq:sigmaveff_1} as
\begin{equation} \label{eq:sigmaveff_2}
\langle \sigma v \rangle_{\rm eff} \lesssim \langle \sigma v \rangle_{\rm obs} \Big(m_{\rm DM} = \xi m_{\hK}\Big)_{X_{\rm SM} \overline{X}_{\rm SM}}, \;\; \langle \sigma v \rangle_{\rm eff} \equiv \frac{\xi^2}{2} \langle \sigma v \rangle_{\hK \hK \to \hpi \he} \left( \frac{ 2 Y_{\hK}^0 m_{\hK} }{Y_{\rm DM} m_{\rm DM}} \right)^2 .
\end{equation}
Note that $\hK \hK\to \hpi \he$ annihilation proceeds through the $s$-wave.

\begin{figure}[t]
    \centering
    \includegraphics[width=0.48\linewidth]{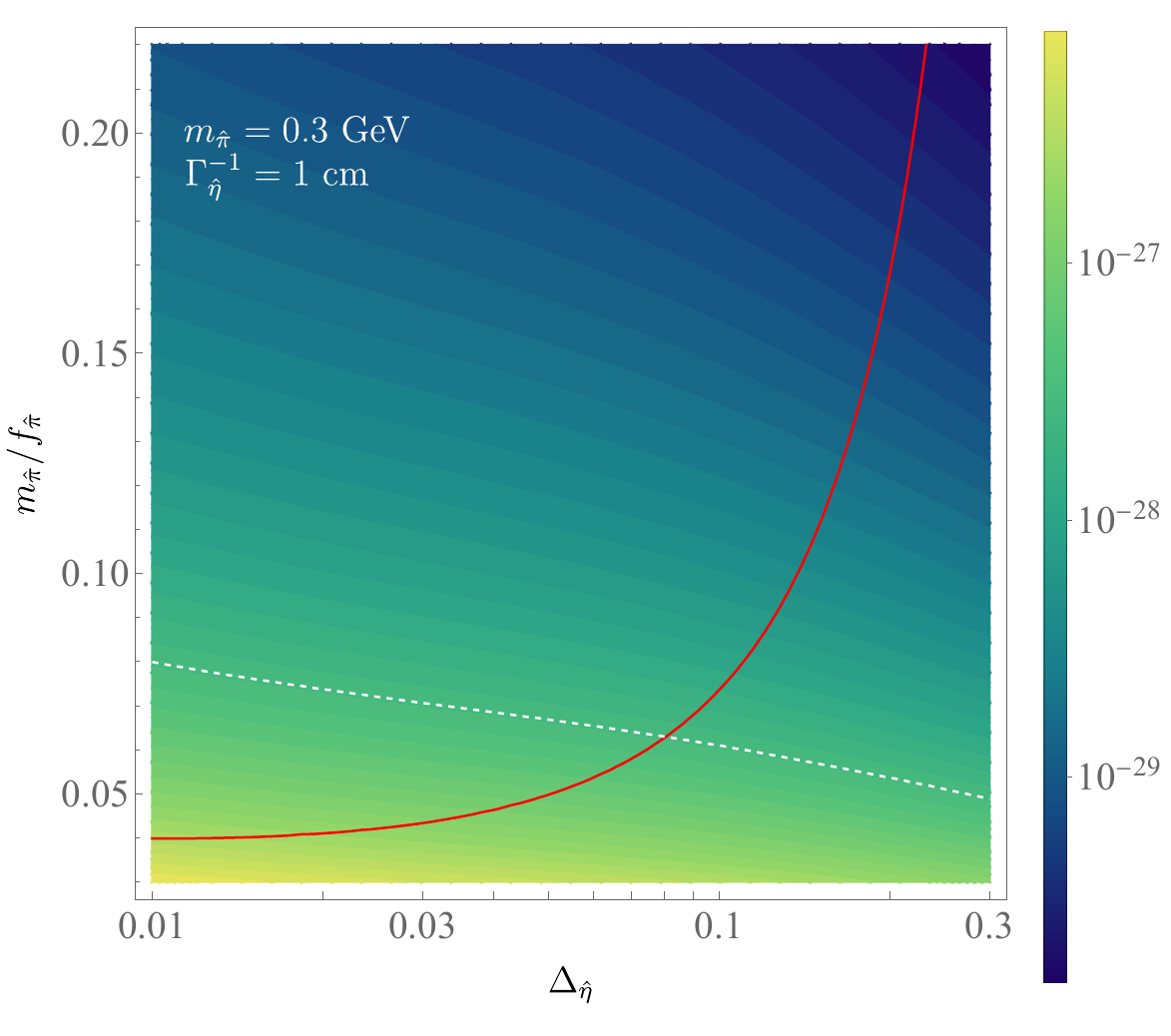}\hspace{2mm}
    \includegraphics[width=0.48\linewidth]{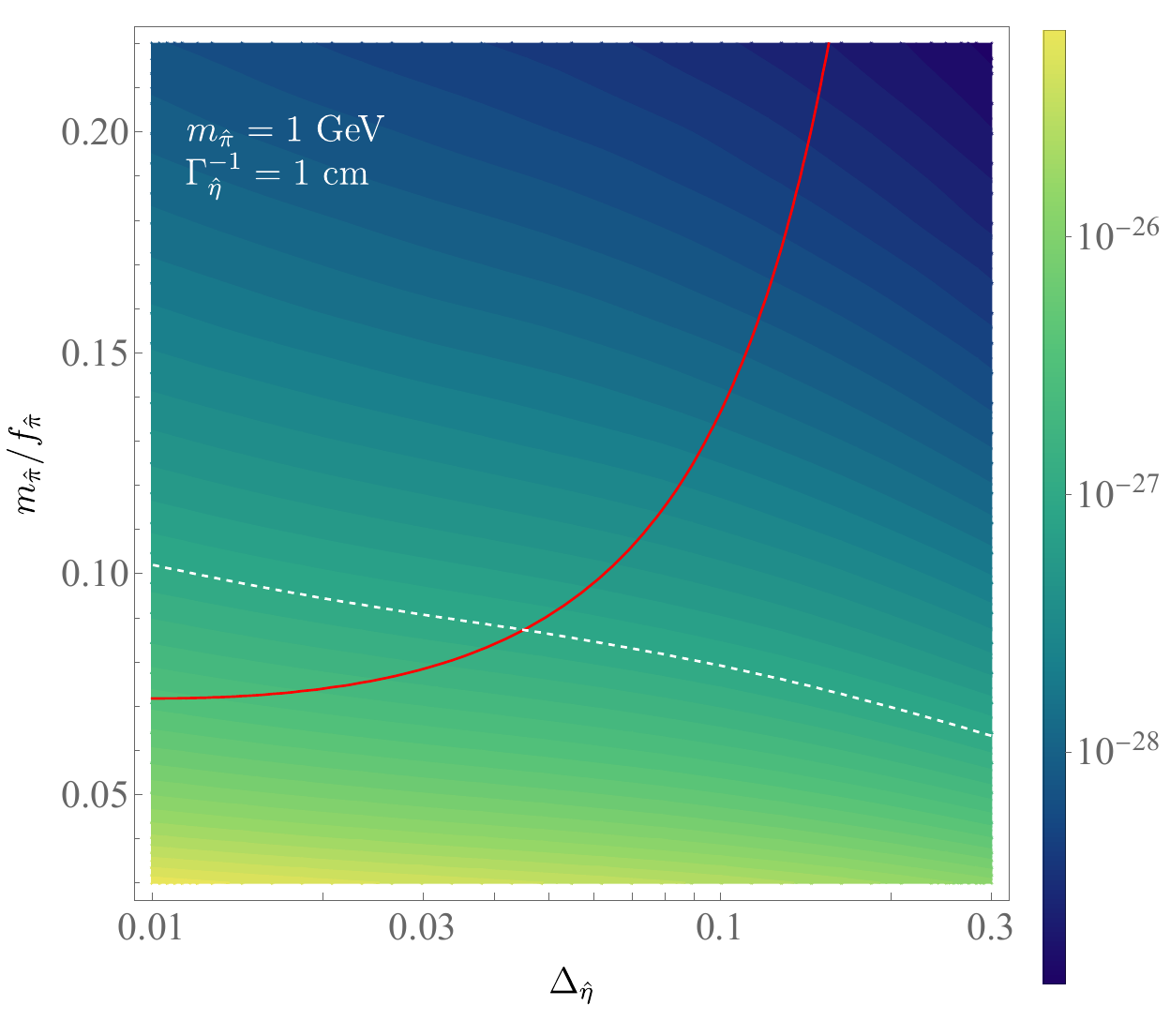}
    \includegraphics[width=0.48\linewidth]{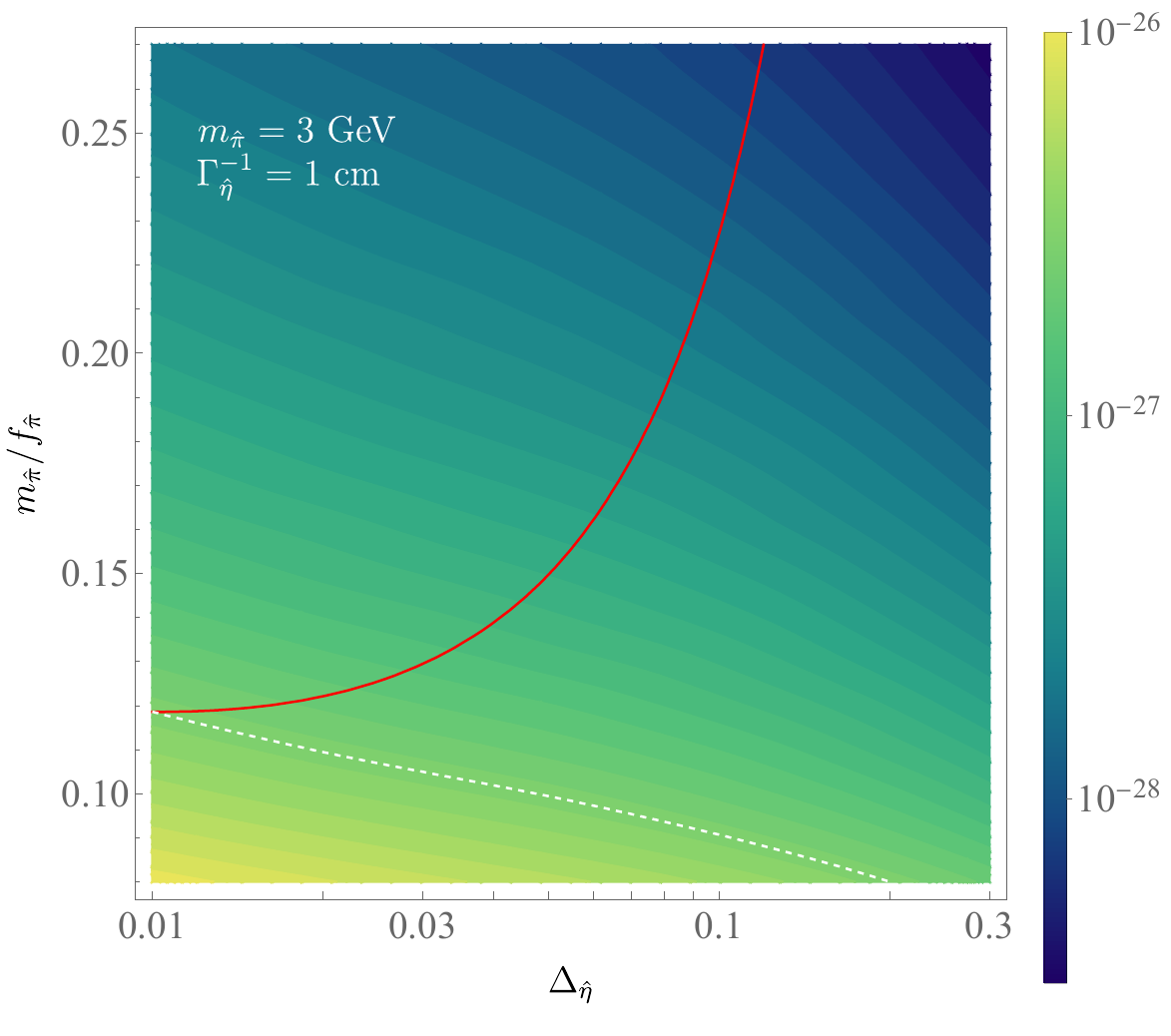}\hspace{2mm}
    \includegraphics[width=0.48\linewidth]{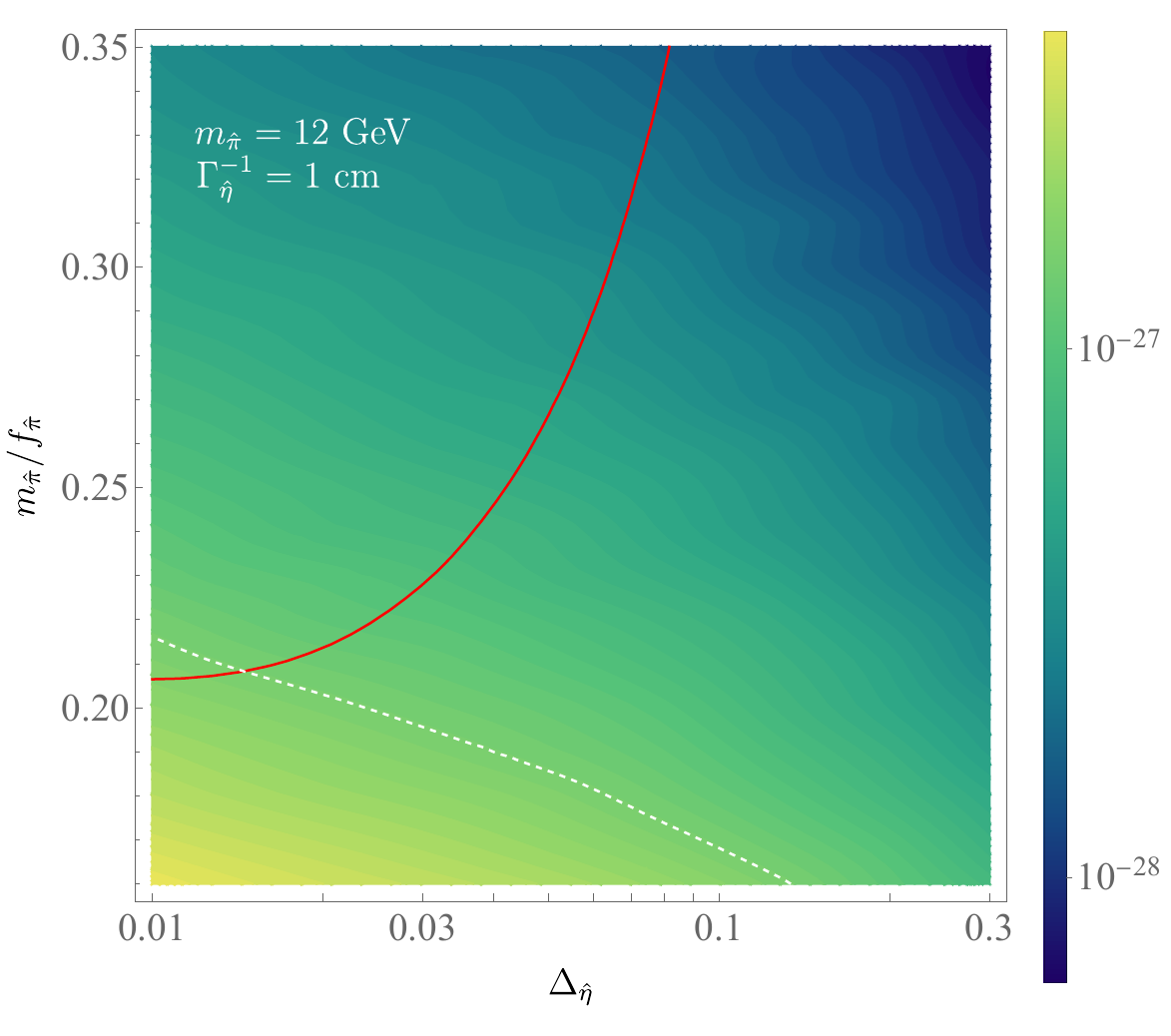}    
    \caption{Contours of the effective annihilation cross section $\langle\sigma v\rangle_{\rm eff}$, in units of cm$^3\;$s$^{-1}$, in the region of small mass splittings. We have fixed $\Gamma_{\he}^{-1}= 1$~cm and the four panels correspond, from upper-left to lower-right, to $m_{\hpi}= \{0.3, 1, 3, 12\}\,$GeV, respectively. The regions below the white dashed contours violate our approximate bound on the $\hK\hK \to \hpi (\he \to \mathrm{SM})$ annihilation process. Along the red contours, the thermal $\hpi+\hK$ abundance matches the observed DM density.}
    \label{fig:degen1}
\end{figure}

A review of the indirect constraints on annihilating DM from astrophysical observations can be found in Ref.~\cite{Cirelli:2024ssz}. In the mass region where the $b\bar{b}$ channel is kinematically open, the strongest constraints for $m_{\rm DM}\geq 6\;\mathrm{GeV}$ have been reported from the combined searches for $\gamma$-rays from dwarf spheroidal galaxies~\cite{Hess:2021cdp}, giving $\langle \sigma v \rangle_{\rm obs} \approx 1.6\times 10^{-27}\, \text{cm}^{3}\,\text{s}^{-1}$ at $m_{\rm DM} = 6\;\mathrm{GeV}$ ($95\%$ CL). Comparable bounds, but reported only for $m_{\rm DM}\geq 10$~GeV, arise from radio searches for synchrotron radiation from the Large Magellanic Cloud~\cite{Regis:2021glv}. At smaller masses, $m_\mu < m_{\rm DM} < 6\;\mathrm{GeV}$, for illustration we can focus on the $\mu^+\mu^-$ channel. The strongest constraints come from the Cosmic Microwave Background anisotropies~\cite{Slatyer:2015jla,Lopez-Honorez:2013cua}, yielding at $95\%$~CL $\langle \sigma v \rangle_{\rm obs} \approx 2\times 10^{-27}\, \text{cm}^{3}\,\text{s}^{-1}\,(m_{\rm DM}/\mathrm{GeV})$~\cite{Lopez-Honorez:2013cua}. Below the $\mu^+\mu^-$ threshold, the $\he$ lifetime would be too long to effectively deplete the dark sector number density in any realistic model, so we do not consider the region $m_{\rm DM} < m_\mu$.

The effective cross section $\langle \sigma v \rangle_{\rm eff}$ defined in Eq.~\eqref{eq:sigmaveff_2} is shown in Fig.~\ref{fig:degen1}, for fixed width $\Gamma_{\he}^{-1} = 1\;\mathrm{cm}$ and four benchmark values of $m_{\hpi}$. Along the red contours, the entirety of DM consists of $\hpi$ and $\hK$. The indirect detection constraints then rule out small values for both $\Delta_{\he}$ and $m_{\hpi} / f_{\hpi}$: for example, for $m_{\hpi}=1$~GeV we find $\Delta_{\he} \simeq 4\hspace{0.2mm} \Delta_{\hK}/3 \gtrsim 0.05$ and $m_{\hpi} / f_{\hpi} \gtrsim 0.09$. Above the red contours, $\hpi$ and $\hK$ do not constitute all of DM, but the indirect detection constraints still apply. The regions below the red contours are instead not viable, as the $\hpi + \hK$ density exceeds the observed value.

The above constraints are affected by two theoretical uncertainties. (1) Our prediction for the present-day $\hK$ number density assumes that the dark and visible sectors shared a common temperature until $\hK$ froze out, but if kinetic decoupling happened earlier, $Y_{\hK}^0$ can be modified by an $\mathcal{O}(1)$ factor (see Ref.~\cite{Katz:2020ywn} for a discussion in a related setup). (2) The pattern of branching ratios of $\he$ to SM particles, which impacts somewhat the sensitivity of observations, is model dependent; in particular, hadronic annihilation channels may compete with or dominate over $\mu^+\mu^-$ at small masses (see Sec.~\ref{sec:UV} for further discussions). However, these effects are not expected to alter the main conclusion we have reached: indirect DM searches rule out mass splittings below a few percent and values of $m_{\hpi}/f_{\hpi}$ smaller than $0.1\,\text{-}\,0.2$, leaving open the rest of the parameter space.

\section{Ultraviolet-Complete Models and Their Signals}
\label{sec:UV}
The signatures at collider experiments depend strongly on the UV completion of the portal that mediates the $\he \to \mathrm{SM}$ decay. Once the UV completion is specified, it also becomes possible to check whether any relevant signals are expected in direct or indirect DM detection, beyond the irreducible (but easily very suppressed) $\hK\hK \to \hpi \he$ process discussed in Sec.~\ref{sec:ID}. As the following discussion demonstrates, our low-energy setup can be UV-completed by several models that were discussed in previous literature on dark showers.

\subsection{$Z$ portal}\label{sec:Z-portal}

In the first UV completion we consider, dark QCD is connected to the SM by introducing heavy dark quarks, which transform in the fundamental representation of $SO(N_d)$ and have charges $\bm{Q} \sim (\mathbf{1},\mathbf{2})_{-1/2}$ and $\bm{Q}^c \sim (\mathbf{1},\mathbf{2})_{+1/2}$ under the $(SU(3)_c, SU(2)_L)_{U(1)_Y}$ gauge group of the SM~\cite{Cheng:2019yai,Cheng:2021kjg}. These quantum numbers admit vector-like masses and Yukawa couplings involving the SM Higgs field,
\begin{equation} 
-\, \mathcal{L}_{\rm UV} \supset Q_i^{c \,T} (M_Q)_{ij} \epsilon Q_{j} + Q_i^T H Y_{ij} \epsilon \psi_j + Q_i^{c\,T} \widetilde{H} \widetilde{Y}_{ij} \epsilon \psi_j + \mathrm{h.c.}. 
\end{equation}
Without loss of generality, we assume that $M_Q$ is diagonal with real and positive entries. Since $U(1)_{T_2}$ acts as a rotation in the $(1,2)$ flavor subspace, requiring that it is an exact symmetry imposes the following structures to the mass and Yukawa matrices,\footnote{To generate a $Z$ portal, it would be sufficient to introduce only $(Q_{1,2},\, Q^c_{1,2})$ or only $(Q_3,\, Q^c_3)$. Here we keep the discussion general.}
\begin{equation}
M_Q = \mathrm{diag}(M_{Q_1}, M_{Q_1}, M_{Q_3})\,,\qquad Y = \mathrm{diag}(y_{1}, y_{1}, y_{3})\,, \qquad \widetilde{Y} = \mathrm{diag}(\tilde{y}_{1}, \tilde{y}_{1}, \tilde{y}_{3})\,.
\end{equation}
In general, the Yukawa matrices contain two physical complex phases, one in each of the (1,2) and (3) flavor subspaces. 

Assuming that $\bm{Q}, \bm{Q}^c$ have masses above the weak scale, they can be integrated out. The resulting tree-level effective Lagrangian is, up to dimension $6$,
\begin{align}
\Delta \mathcal{L}_{\rm EFT}^{\bm{Q}, \bm{Q}^c}  =\;& \frac{1}{2} \bigg[ \frac{|y_1|^2 - |\tilde{y}_1|^2}{M_{Q_1}^2}\, \big(\psi_{1}^\dagger \bar{\sigma}^\mu \psi_{1} + \psi_{2}^\dagger \bar{\sigma}^\mu \psi_{2} \big) + \frac{|y_3|^2 - |\tilde{y}_3|^2}{M_{Q_3}^2}\, \psi_{3}^\dagger \bar{\sigma}^\mu \psi_{3}  \bigg] (i H^\dagger D_\mu H + \mathrm{h.c.} ) \nonumber \\
+\;&\bigg[  \frac{ y_1 \tilde{y}_1}{M_{Q_1}}  \big(\psi_1^T \epsilon \psi_1 + \psi_2^T \epsilon \psi_2\big) + \frac{y_3 \tilde{y}_3}{M_{Q_3}} \psi_3^T \epsilon \psi_3  \bigg] |H|^2 + \mathrm{h.c.} \label{eq:EFT_total} \\
+\;& \frac{1}{2}\bigg[  \frac{|y_1|^2 + |\tilde{y}_1|^2}{ M_{Q_1}^2}\, m_1   \big(\psi_1^T \epsilon \psi_1 + \psi_2^T \epsilon \psi_2\big) + \frac{|y_3|^2 + |\tilde{y}_3|^2}{ M_{Q_3}^2}\, m_3  \psi_3^T \epsilon \psi_3  \bigg] |H|^2 + \mathrm{h.c.}. \nonumber  
\end{align}
In the third line, the equations of motion for the $\bm{\psi}$ fields arising from Eq.~\eqref{eq:L_psi} were used.

The first line of Eq.~\eqref{eq:EFT_total} contains dimension-$6$ operators that induce couplings of the dark quarks to the $Z$ boson, as long as $|y_i|\neq |\tilde{y_i}|$,
\begin{equation}\label{eq:A_def}
 - \frac{g_Z}{2} {\bm{\psi}^\dagger} \bar{\sigma}^\mu \mathcal{A} \bm{\psi} Z_\mu\, ,\qquad \mathcal{A} \equiv \frac{v^2}{2} \mathrm{diag}\hspace{0.3mm}\bigg( \frac{|y_1|^2 \,-\, |\tilde{y}_1|^2}{M_{Q_1}^2} ,\, \frac{|y_1|^2 \,-\, |\tilde{y}_1|^2}{M_{Q_1}^2} ,\, \frac{|y_3|^2 \,-\, |\tilde{y}_3|^2}{M_{Q_3}^2}\bigg) \,,
\end{equation}
where $v \approx 246\;\mathrm{GeV}$. The couplings of the dark mesons to the $Z$ boson are obtained by making the replacement $\partial_\mu \Sigma \to D_\mu \Sigma = \partial_\mu \Sigma + i g_Z Z_\mu ( \mathcal{A} \Sigma + \Sigma \mathcal{A}^T )/2$ in the chiral Lagrangian, Eq.~\eqref{eq:p2}. Focusing on terms that are linear in $Z_\mu$, we find
\begin{align} \label{eq:single_Z_terms}
\mathcal{L}_{\hpi,\,p^2} \supset&\; g_Z f_{\hpi}   \mathrm{Tr} (X_i \mathcal{A}) Z_\mu \partial^\mu \hpi_i  
%+ {\color{red}\,0\, \times\,} Z_\mu \hpi \partial^\mu \hpi 
 + \mathcal{O}(Z_\mu \hpi^2 \partial^\mu \hpi) =  \frac{2\hspace{0.2mm}m_Z^2}{g_Z F_{\he}}\hspace{0.3mm}  Z_\mu  \partial^\mu \he_0 +  \mathcal{O}(Z_\mu \hpi^2 \partial^\mu \hpi ) \,,
\end{align}
where the trace that determines the linear $\hpi_i\,$-$\,Z$ mixing is nonzero only for $i = 8$, hence we have defined the (inverse of the) effective ALP decay constant for the unstable $\he$,
\begin{equation}
\frac{\mathrm{PeV}}{F_{\he}} = \frac{1}{\sqrt{3}}\,  \frac{f_{\hpi}}{\mathrm{GeV}}  \bigg[ \frac{|y_1|^2 - |\tilde{y}_1|^2}{(M_{Q_1}/\mathrm{TeV})^2} - \frac{|y_3|^2 - |\tilde{y}_3|^2}{(M_{Q_3}/\mathrm{TeV})^2} \bigg]\,.
\end{equation}
Notice that the $Z$ couplings to an even number of dark mesons, including the $\mathcal{O}(Z_\mu \hpi \partial^\mu \hpi)$ term that could mediate scattering of DM on SM particles, vanish exactly due to $\mathcal{A} = \mathcal{A}^\ast$. The absence of the $\mathcal{O}(Z_\mu \hpi \partial^\mu \hpi)$ term is a consequence of the dark charge conjugation symmetry $\mathcal{C}$ discussed in Sec.~\ref{sec:portal}. At energies below $m_Z$, the $Z$ boson can be integrated out and the first term in Eq.~\eqref{eq:single_Z_terms} yields dimension-$5$ interactions of $\he$ with the SM fermions,
\begin{equation} \label{eq:Z_portal}
\mathcal{L}_{\he} = -\, \frac{\partial_\mu \he_0}{F_{\he}} \sum_{f\,\in\, q,\, l} c_f  \bar{f} \gamma^\mu \gamma_5 f\,, \qquad c_f = T_{Lf}^3 \,,
\end{equation}
hence $c_u = +1/2$, $c_d = -1/2$, and so on for all the other quarks and leptons. 

The second and third lines of the EFT Lagrangian in Eq.~\eqref{eq:EFT_total} contain operators that generate seesaw-like masses and couplings to the Higgs boson for the light dark quarks. In general, they lead to DM-nucleon scattering mediated by Higgs exchange, which is allowed by $\mathcal{C}$-invariance. We find that this signal is marginally accessible for the operators in the second line, which however can be naturally suppressed, if the Yukawa couplings $y_i$ and $\tilde{y}_i$ are hierarchical due to (approximate) chiral symmetries acting on the dark quarks, \eg~$|\tilde{y}_i| \ll |y_i|$. The operators in the third line survive even in the limit $\tilde{y}_i \to 0$, but we find that the corresponding DM-nucleon scattering cross section sits out of current experimental reach, well inside the neutrino fog. The effects of these Higgs portal operators are discussed in Appendix~\ref{app:Higgs}, for completeness.

\subsubsection{Current and projected collider sensitivity}
\label{ssec:Zportal}
The $Z$-portal UV completion leads to rich collider phenomenology. The effective interaction in Eq.~\eqref{eq:A_def} mediates $Z$ decays to dark quarks, with branching ratio~\cite{Cheng:2021kjg}
\begin{equation}
    \text{BR}_Z \big(\bm{\psi}\overline{\bm{\psi}}\hspace{0.2mm}\big) = \frac{N_d m_Z^3 \text{Tr}(\mathcal{A}^\dagger \mathcal{A})}{24 \pi v^2\hspace{0.3mm}\Gamma_Z}  \approx 6\times 10^{-4}\,  \frac{N_d}{5} \bigg[ \frac{ (|y_1|^2 - |\tilde{y}_1|^2)^2}{(M_{Q_1}/\mathrm{TeV})^4} +  \frac{(|y_3|^2 - |\tilde{y}_3|^2)^2}{2 (M_{Q_3}/\mathrm{TeV})^4} \bigg]\,.
    %}
\end{equation}
If the dark pNGBs are sufficiently light, $m \ll 10$~GeV, the dark quarks undergo parton shower and hadronization dynamics analogous to SM QCD, forming dark jets that mainly contain pNGB mesons. We estimate that $\he$ particles, which are the only dark jet components to decay back to the SM, account for $\lesssim 1/5$ of the multiplicity for the moderate mass splittings we consider. Due to the large number of $Z$ bosons produced at the LHC, these dark shower signals can be constrained by inclusive searches for light LLPs. Dark mesons can also be produced through flavor-changing neutral-current (FCNC) decays of SM $B, D$ and $K$ mesons, if kinematically allowed. These processes are mediated by electroweak loop diagrams. Focusing on $B$ mesons, the exclusive rates for FCNC decays to $\he$ are~\cite{Cheng:2021kjg,Cheng:2024hvq}
\begin{equation}
    \text{BR}_B\big(\{K,K^{\ast}\} +\he \big) \approx 10^{-8}\,\bigg(\frac{\text{PeV}}{F_{\he}}\bigg)^2 \bigg(\frac{\mathcal{K}_t}{10}\bigg)^2 \big\{\lambda_{BK\he}^{1/2}, \lambda_{BK^\ast\he}^{3/2}\big\}\,,
\end{equation}
where $\mathcal{K}_t$ contains a logarithmic dependence $\sim \log\, \small( M_{Q_i}^2/m_t^2 \small)$ on the heavy dark quark masses, and $M_{Q_i}\sim 1\;\mathrm{TeV}$ has been taken as a reference. Here $\lambda_{ijk} \equiv \lambda[1, m_j^2/m_i^2, m_k^2/m_i^2]$ encodes the two-body decay kinematics, where $\lambda$ is defined in Eq.~\eqref{eq:lambda_def}.

\begin{figure}
\label{fig:BR_1}
    \centering
    \includegraphics[width=0.7\linewidth]{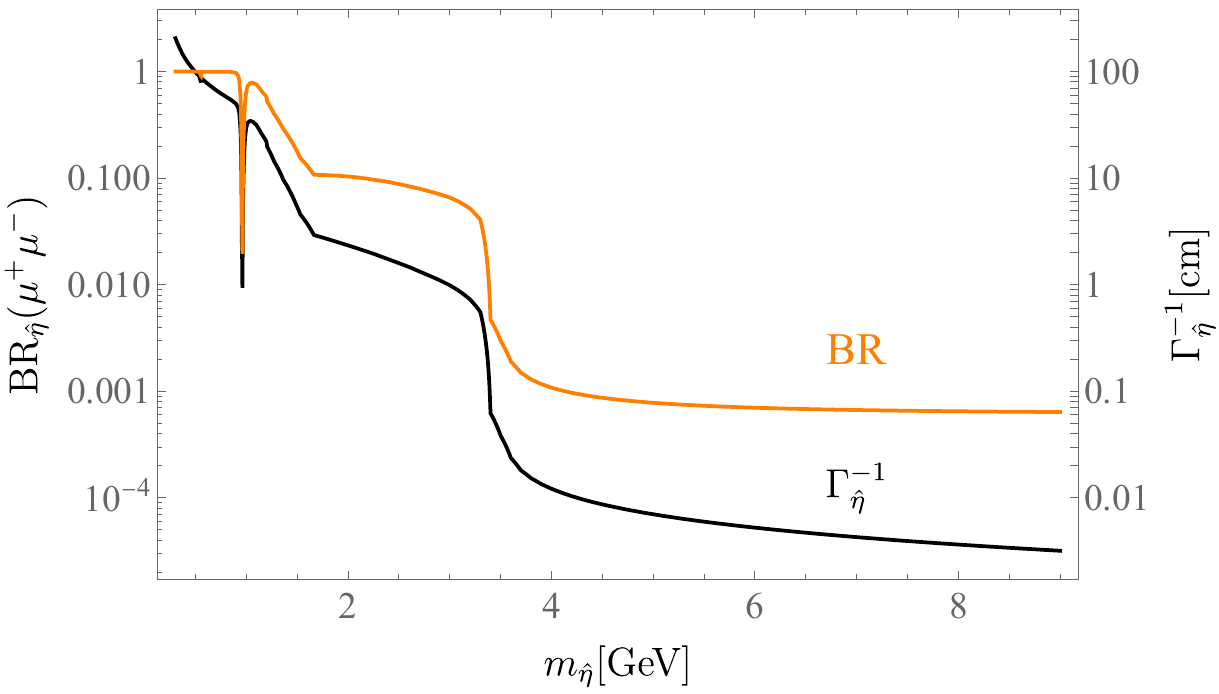}
    \caption{The branching ratio to $\mu^+ \mu^-$ (orange) and inverse proper decay length (black) as functions of the mass, for a pseudoscalar $\he$ coupled to SM fermions through the $Z$ portal, see Eq.~\eqref{eq:Z_portal}. The decay length is calculated for a reference decay constant $F_{\he} = $ 1~PeV.\label{fig:eta_BR}}
\end{figure}

The decays to SM particles of a GeV-scale pseudoscalar coupled through the $Z$ portal in Eq.~\eqref{eq:Z_portal} were first studied in Ref.~\cite{Cheng:2019yai}, extending earlier results~\cite{Aloni:2018vki} (see also later analyses in Refs.~\cite{DallaValleGarcia:2023xhh,Ovchynnikov:2025gpx,Bai:2025fvl}). The calculation of exclusive hadronic branching ratios has been recently improved with manifestly field-basis-independent results in Ref.~\cite{Balkin:2025enj}. Field-basis independence requires additional Lagrangian terms which were missed in previous studies, and that significantly modify the widths of some exclusive hadronic modes. In this paper we adopt the results of Ref.~\cite{Balkin:2025enj} for hadronic exclusive decays, while for the $\gamma\gamma$ and $\ell^+ \ell^-$ ($\ell = e, \mu, \tau$) final states we follow Ref.~\cite{Cheng:2019yai}.\footnote{In Ref.~\cite{Balkin:2025enj}, the couplings to charm and heavier quarks were ignored. Here we include the charm loop contribution to the perturbative $\he \to gg$ decay, which together with $\he \to q\bar{q}$ makes up the total hadronic width at larger mass. The matching to the total hadronic width at smaller mass, which is the sum of the exclusive hadronic modes presented in Ref.~\cite{Balkin:2025enj}, is made at $m_{\he} \approx 1.7$~GeV.} The resulting branching ratio of $\he$ to the dimuon final state, which is the most striking one from an experimental perspective, is shown in Fig.~\ref{fig:eta_BR}. We observe that $\he$ decays dominantly to $\mu^+\mu^-$ if $2 m_\mu < m_{\he} < m_{\eta^\prime} \approx 0.96$~GeV, due to suppression of the $3\pi$ modes. Below $2m_\mu$, $\he$ decays to $e^+e^-$ and $\gamma\gamma$, but the lifetime is too long for the slide DM mechanism to be effective. Above $m_{\eta^\prime}$, mixing with SM pseudoscalar mesons leads to a variety of hadronic final states, with large branching ratios to $\eta\pi\pi$ and $\pi K K$. In Fig.~\ref{fig:eta_BR} we show the proper decay length of $\he$, as well, taking $F_{\he}=1$~PeV as reference value for the effective ALP decay constant.

$Z$ decays and FCNC $B$ meson decays produce relatively soft $\he$'s, without additional hard objects that can be exploited to trigger on these events, posing challenges to searches at the LHC. Specialized analysis strategies and/or detectors that target displaced $\he$ decays are required. So far, the data scouting technique applied by CMS~\cite{CMS:2021sch} and the LHCb Vertex Locator detector~\cite{LHCb:2020ysn} were shown to be sensitive to $\he \to \mu^+\mu^-$ decays~\cite{Cheng:2021kjg,Cheng:2024hvq,Cheng:2024aco}. Furthermore, auxiliary detectors located far away from the LHC interaction points can be useful to search for $\he$'s with longer decay lengths. We obtain current LHC constraints by recasting the results of inclusive searches for LLPs decaying to dimuons at CMS~\cite{CMS:2021sch} and LHCb~\cite{LHCb:2020ysn}. To project the future experimental sensitivity, we follow the approach in Ref.~\cite{Cheng:2024aco}, including searches for individual dimuon displaced vertices (DVs) at multipurpose detectors and inclusive searches for displaced decays at far auxiliary detectors.

\begin{figure}
    \centering
    \includegraphics[width=0.47\linewidth]{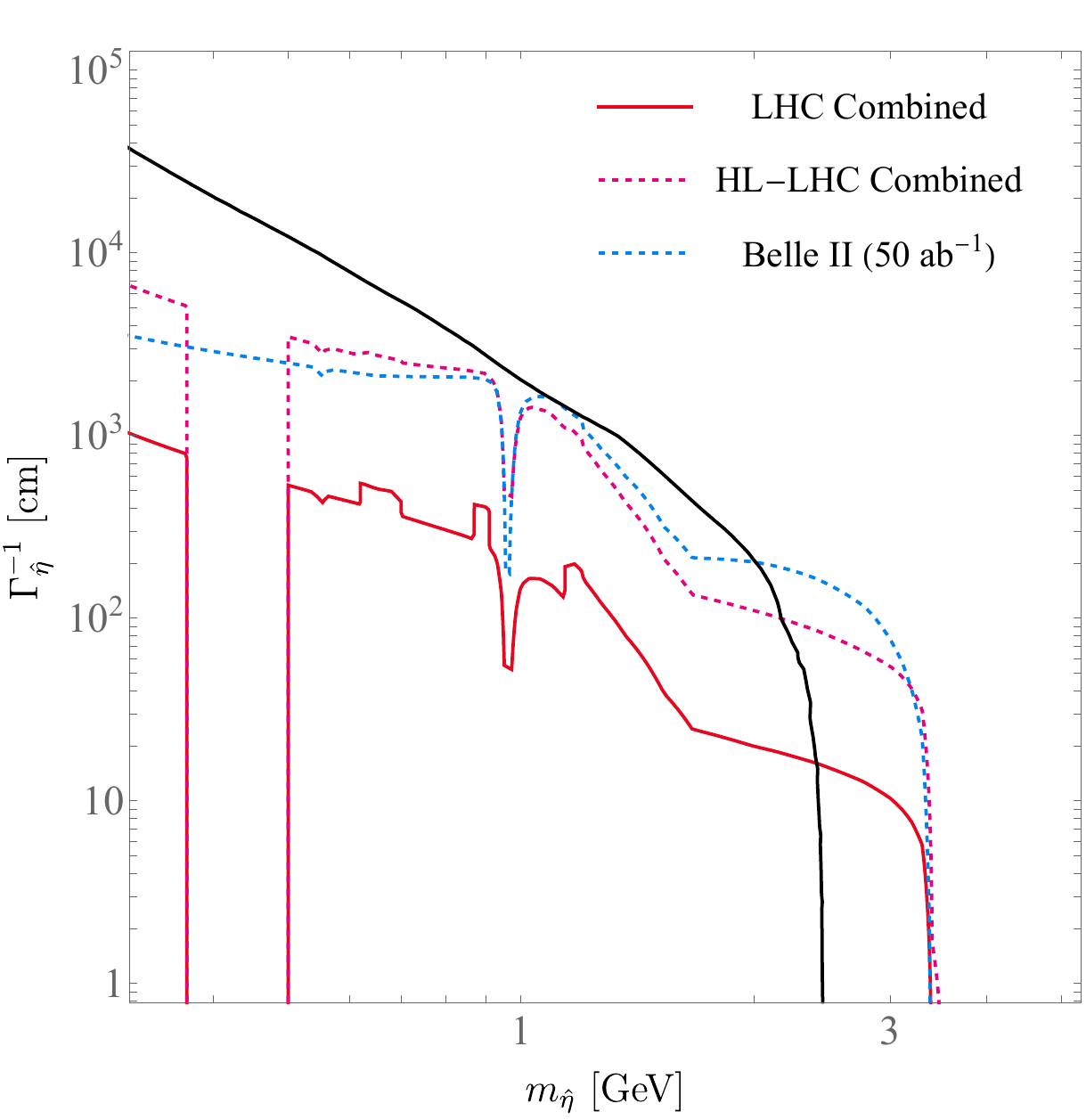}\hspace{4mm}
    \includegraphics[width=0.47\linewidth]{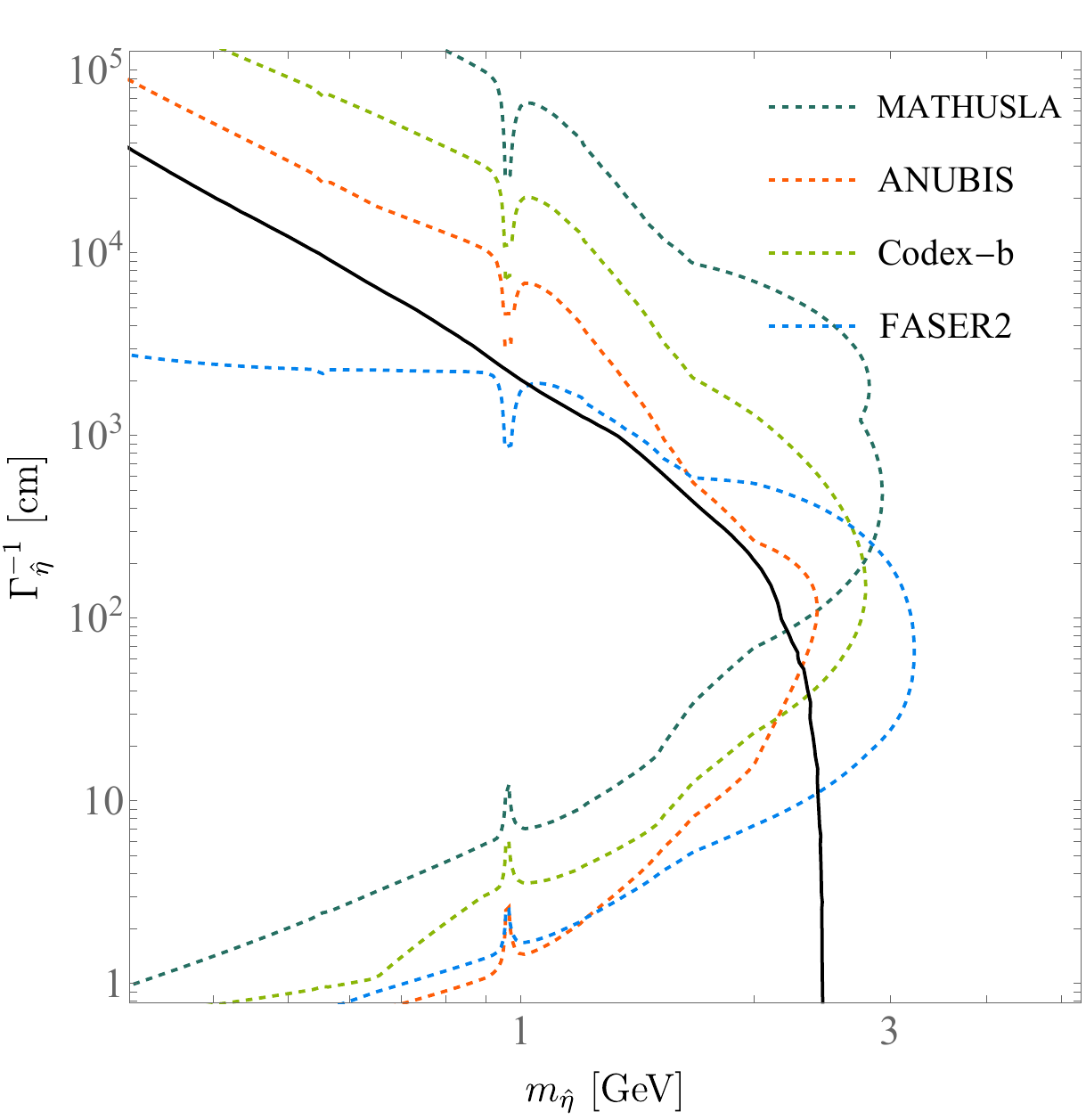}    
    \caption{Current and projected collider constraints on a benchmark $Z$-portal model with $N_d = 5$, $m_{\hpi}/f_{\hpi} = 1$ and $\Delta_{\he}=0.3$, obtained from multipurpose detectors~(left) and auxiliary detectors~(right). Along the solid black contour, $\hpi_{\pm}$ make up the observed DM relic density. Where applicable, the projections combine dark shower signals initiated by $Z$ decays and inclusive FCNC decays of $B$ mesons.\label{fig:pheno1}}
\end{figure}

To report our results, we select a representative benchmark model where
\begin{equation}
N_d = 5\,,\quad m_{\hpi}/f_{\hpi}=1\,,\quad \Delta_{\he}=0.3\,,\qquad M_{Q_3}\to \infty\,,\quad \tilde{y}_1 = 0\,.
\end{equation}
With these choices, the UV Lagrangian is automatically $CP$ conserving. Picking a pair $(m_{\he}, \Gamma_{\he})$ determines the value of $y_1^2/M_{Q_1}^2$, which in turn fixes the overall size of the effective coupling matrix $\mathcal{A}$. The left panel of Fig.~\ref{fig:pheno1} shows the combined LHC sensitivity of searches for $Z$-initiated dark shower signals and inclusive $B\to X_s (\he\to \mu^+\mu^-)$ decays at multipurpose detectors, following the single dimuon DV search approach described in Ref.~\cite{Cheng:2024aco}. High-Luminosity LHC (HL-LHC) limits are also projected, including higher statistics and estimated improvements in experimental systematics. Moreover, the projected constraints from a search for inclusive FCNC $B$ decays in $50$~ab$^{-1}$ of Belle II data are presented. For the assumed benchmark model, multipurpose detectors will cover the large decay width region up to $\Gamma^{-1}_{\he}\sim \mathcal{O}(1)\;\mathrm{m}$. The right panel of Fig.~\ref{fig:pheno1} highlights the complementary role of future auxiliary detectors in probing the small decay width region. The projected limits from the ANUBIS~\cite{Bauer:2019vqk,ANUBIS:2025sgg}, Codex-b~\cite{CODEX-b:2025rck}, FASER2~\cite{FASER:2018eoc} and MATHUSLA~\cite{MATHUSLA:2025zyt} detectors are shown, obtained by assuming that all visible $\he$ decays within the detector volume can be detected with $\mathcal{O}(1)$ efficiency. 
%\hc{except ANUBIS?} 
The displayed contours combine the $\he$ signals from inclusive FCNC $B$ decays and $Z$-initiated dark showers where relevant. Further details on our projections can be found in Appendix~\ref{app:phenomenology}.
In all cases, the constraints weaken for $m_{\he}\gtrsim 3$~GeV, where the $\he\to c\bar{c}$ decay is kinematically open: as the $c\bar{c}$ channel dominates the $\he$ decay width, a much smaller effective coupling $y_1^2/M_{Q_1}^2$ is needed to keep $\he$ long lived (and therefore detectable by the searches we consider). In turn, this leads to much smaller production rates. In addition, $\mathrm{BR}_{\hspace{0.2mm}\he}(\mu^+\mu^-)$ becomes highly suppressed, further reducing the sensitivity of dimuon DV searches. 

\subsection{$Z'$ portal}
\label{sec:zprime}
The second UV completion we consider connects dark QCD and the SM through a $Z'$ vector boson, associated to a $U(1)^\prime$ gauge group. The interactions between the dark quarks and the $Z^\prime$ are
\begin{align}
  \mathcal{L}_{\rm UV} \supset  -\,g_d\hspace{0.3mm} {\bm{\psi}}^\dagger\bar{\sigma}^\mu \mathcal{Q}\hspace{0.3mm} \bm{\psi} Z'_\mu  \,,
\end{align}
where $g_d$ is the $U(1)^\prime$ gauge coupling and the charge matrix has the structure $\mathcal{Q} = \mathrm{diag}\hspace{0.3mm}(q_1, q_1, q_3)$ to preserve the $U(1)_{T_2}$ symmetry that stabilizes DM. The couplings of the dark mesons to the $Z^\prime$ are obtained through the replacement $\partial_\mu \Sigma \to D_\mu \Sigma = \partial_\mu \Sigma + i g_d Z^\prime_\mu ( \mathcal{Q} \Sigma + \Sigma \mathcal{Q}^T )$ in the chiral Lagrangian of Eq.~\eqref{eq:p2}. The linear terms in $Z^\prime_\mu$ are found to be
\begin{align} \label{eq:single_Zp_terms}
\mathcal{L}_{\hpi,\,p^2} \supset&\; 2 g_d  f_{\hpi}  \mathrm{Tr} (X_i \mathcal{Q}) Z^\prime_\mu  \partial^\mu \hpi_i  + \mathcal{O}(Z^\prime_\mu \hpi^2 \partial^\mu \hpi) =  \frac{2 M_{Z'}^2}{g_d F_{\he}}\hspace{0.3mm}  Z^\prime_{\mu}  \partial^\mu \he_0 +  \mathcal{O}(Z^\prime_\mu \hpi^2 \partial^\mu \hpi ) \,,
\end{align}
where the effective ALP decay constant was defined as
\begin{equation}
\frac{\mathrm{PeV}}{F_{\he}} = \frac{1}{\sqrt{3}}\, \frac{f_{\hpi}}{\mathrm{GeV}}\, \frac{g_d^2\, (q_1-q_3)}{(M_{Z'}/\mathrm{TeV})^2}\, .
\end{equation}
The $\mathcal{O}(Z_\mu \hpi \partial^\mu \hpi)$ term in Eq.~\eqref{eq:single_Zp_terms}, which could couple the $Z'$ to the DM vector current, vanishes due to the $\mathcal{C}$ symmetry.

The phenomenology depends on how the $Z^\prime$ interacts with the SM fields. If at least some SM fermions are charged under $U(1)^\prime$, the $Z'$ needs to be heavier than the weak scale and can be integrated out, yielding couplings of $\he$ to the axial-vector current built out of SM fermions,
\begin{equation}
\mathcal{L}_{\he} = -\, \frac{\partial_\mu \he_0}{F_{\he}} \sum_{f\,\in\, q,\, l} c_f  \bar{f} \gamma^\mu \gamma_5 f\,, \qquad c_f = q_{f_L}-q_{f_R} \,.
\end{equation}
Hence, the $U(1)^\prime$ charges of some SM fermions need to be chiral for this portal to be effective\footnote{If all SM fermions have vector-like charges under $U(1)^\prime$, but the latter is anomalous with respect to the SM electroweak group (\eg baryon number), then $\he$ has effective couplings to pairs of electroweak gauge bosons ($WW, ZZ$ and $Z\gamma$). However, in this case the resulting lifetime will be much longer, rendering the slide DM mechanism not viable. If all SM fermions are neutral under $U(1)^\prime$, but $Z$-$Z'$ mass mixing is induced by an additional Higgs doublet carrying $U(1)^\prime$ charge, then the effective couplings of $\he$ have the structure obtained in Sec.~\ref{sec:Z-portal}, see Eq.~\eqref{eq:Z_portal}. In this case the $Z'$ could be much lighter, with interesting phenomenology~\cite{Cheng:2024hvq,Cheng:2024aco}.} (see,~\eg Ref.~\cite{ParticleDataGroup:2024cfk} for a review of $Z'$ models). The gauge anomalies can be canceled by additional fermions around the $Z'$ mass scale. To avoid potentially dangerous FCNC effects,
it may be preferable to assume generation-independent charges $q_{f_L}$ and $q_{f_R}$. For a given pattern of charges, the decays of $\he$ can be evaluated using the results of Ref.~\cite{Balkin:2025enj}.

Assuming the $Z'$ couples to SM quarks, LHC signatures are dominated by \mbox{$s$-channel} $Z^\prime$ production and decay to dark jets. If $\he$ decays promptly, the dark jets are semi-visible~\cite{Cohen:2015toa}. So far, LHC searches have focused on scenarios where $\he$ decays hadronically. The results of Ref.~\cite{CMS:2021dzg} show that the experimental bound on the $Z^\prime$ production rate is highly dependent on $M_{Z^\prime}$ and the fraction of invisible dark hadrons, $r_{\rm inv}$, but not on the decaying dark hadron mass. 
The constraints become weaker for large $r_{\rm inv}$. In our setup, where $r_{\rm inv}\gtrsim 0.8$, $\sigma(pp\to Z^\prime \to \psi\overline{\psi})$ is constrained to be $\lesssim 10\,(0.1)$~pb for $M_{Z^\prime} =2\,(5)$~TeV. A similar search was carried out in Ref.~\cite{ATLAS:2025kuz}, for $r_{\rm inv}\lesssim 0.6$. 

As the $\he$ lifetime increases, the dark jets become emerging~\cite{Schwaller:2015gea}. Again, current LHC searches focus on hadronically decaying dark mesons. For the optimal dark meson lifetime $\sim 1$~cm, the $\sigma(pp\to Z^\prime \to \psi\overline{\psi})$ limit reaches $\mathcal{O}(1)\,\mathrm{fb}$ for $M_{Z'} = 2\;\mathrm{TeV}$~\cite{ATLAS:2025bsz}. However, that search is not directly applicable here, as it assumed that all dark hadrons decay. Instead, our scenario generically predicts emerging jets that are also semi-visible, as recently considered in Ref.~\cite{Carrasco:2025bct}. In that work, a modified analysis strategy based on the ATLAS emerging jet search~\cite{ATLAS:2025bsz} was optimized for dark meson lifetimes of $\mathcal{O}(10)$~mm. In addition, searches using energy deposits in the CMS muon system~\cite{CMS:2021juv} and ATLAS hadronic/electromagnetic calorimeters~\cite{ATLAS:2024ocv} were recast, covering lifetimes up to $10^3$~cm and $r_{\rm inv}\sim 0.8$. The optimal bound on $\sigma(pp\to Z^\prime \to \psi\overline{\psi})$ can reach down to $\lesssim 0.02$~fb when $M_{Z^\prime} =2$~TeV and $m_{\he} \gtrsim 5$~GeV~\cite{Carrasco:2025bct}. If the $\he$ lifetime is even longer, auxiliary detectors will be capable of probing its decays. If all dark hadrons are effectively stable on collider length scales, the leading constraints arise from mono-X searches: mono-jet searches have placed a $\sim 2$ TeV bound on $M_{Z'}$, assuming a coupling to SM quarks equal to $0.25$~\cite{CMS:2021far,ATLAS:2021kxv}. Additional, model-dependent aspects include the production of $\he$ in FCNC decays of $B$ or $K$ mesons, where kinematically allowed, and the effects of possible couplings of the $Z^\prime$ and $\he$ to leptons.

\subsection{Bi-fundamental scalar portal}\label{sec:bifund}
A third possibility to connect dark QCD with the SM is to introduce scalar fields $\bm{X}$ that transform under both dark color and the SM gauge group, such that Yukawa interactions of the form
\begin{align}
\label{eq:bifundamental}
 \kappa_{ij,\alpha}\hspace{0.2mm} {\psi}^\dagger_i  X_{j}^\ast  f_\alpha + \mathrm{h.c.}
\end{align}
can be written down, where $f_\alpha$ is a chiral SM fermion~\cite{Bai:2013xga,Schwaller:2015gea,Beauchesne:2017yhh,Renner:2018fhh,Carmona:2024tkg}. The $\bm{X}$ scalars transform in the fundamental representation of $SO(N_d)$, while their SM quantum numbers are determined by those of $f_\alpha\hspace{0.2mm}$: for example, $\bm{X} \sim (\bm{3}, \bm{1})_{-1/3}$ under $(SU(3)_c, SU(2)_L)_{U(1)_Y}$ if $f_\alpha$ is a right-handed down-type quark. If a single $X_{j}$ couples to multiple generations of SM fermions, in general FCNCs will be induced, leading to important constraints from flavor physics observables. For illustration, here we consider a model where only a single SM fermion (taken to be the right-handed down quark, $d_R$) is involved in the Yukawa interactions of Eq.~\eqref{eq:bifundamental}. This flavor choice has previously bean adopted in LHC searches~\cite{CMS:2018bvr,CMS:2024gxp}. The Lagrangian contains the terms
\begin{align}
  -\, \mathcal{L}_{\rm UV} \supset   \kappa_{ij}\hspace{0.2mm} {\psi}^\dagger_i X_{j}^\ast  d_R + \mathrm{h.c.} + X_i^\ast (M^2_X)_{ij} X_j\,,
\end{align}
where the flavor structures of the Yukawa and mass matrices are assumed to be
\begin{equation}
\kappa = \mathrm{diag}(\kappa_1, \kappa_1, \kappa_3)\,,\qquad M_X^2 = \mathrm{diag}\big(M_{X_1}^2, M_{X_1}^2, M_{X_3}^2\big)\,,
\end{equation}
to preserve the $U(1)_{T_2}$ symmetry that stabilizes DM. The couplings $\kappa_1$ and $\kappa_3$ can be made real without loss of generality.

The $\bm{X}$ scalars, which carry SM color charges, must have masses at or above the TeV scale to satisfy phenomenological constraints. Integrating them out at tree level induces four-fermion dimension-$6$ operators,
\begin{align}
     \Delta \mathcal{L}_{\rm EFT}^{\bm{X}} =&\;\frac{\kappa_1^2}{M^2_{X_1}} \Big( {\psi}^\dagger_1 d_R\,  {d}_R^{\dagger} \psi_1 +  {\psi}^\dagger_2 d_R\,  {d}_R^{\dagger} \psi_2 \Big) +  \frac{\kappa_3^2}{M^2_{X_3}}\, {\psi}^\dagger_3 d_R\,  {d}_R^{\dagger} \psi_3   \nonumber  \\
     =& - \frac{1}{2} {\bm{\psi}^\dagger} \bar{\sigma}^\mu \mathcal{K}\bm{\psi} \, {d}^\dagger_R \sigma_\mu d_R\,,\qquad \mathcal{K} \equiv \mathrm{diag}\,\bigg( \frac{\kappa_1^2}{M_{X_1}^2},\,  \frac{\kappa_1^2}{M_{X_1}^2},\, \frac{\kappa_3^2}{M_{X_3}^2} \bigg)\,,
\end{align}
where a Fierz transformation was performed to obtain the second line, and $\sigma^\mu \equiv (1, \sigma^i)$.

The couplings of the dark mesons to the SM quark current $J_\mu \equiv d_R^\dagger \sigma_\mu d_R$ are obtained by making the replacement $\partial_\mu \Sigma \to D_\mu \Sigma = \partial_\mu \Sigma + i J_\mu ( \mathcal{K} \Sigma + \Sigma \mathcal{K}^T )/2$ in the chiral Lagrangian of Eq.~\eqref{eq:p2}. Focusing on linear terms in $J_\mu$, we find
\begin{align} \label{eq:single_J_terms}
\mathcal{L}_{\hpi,\,p^2} \supset&\;  f_{\hpi}  \mathrm{Tr} (X_i \mathcal{K}) J_\mu  \partial^\mu \hpi_i  + \mathcal{O}(J_\mu \hpi^2 \partial^\mu \hpi) =  \frac{1}{F_{\he}}  J_\mu  \partial^\mu \he_0 +  \mathcal{O}(J_\mu \hpi^2 \partial^\mu \hpi ) \,,
\end{align}
where the effective ALP decay constant is
\begin{equation}
    \frac{\mathrm{PeV}}{F_{\he}} = \frac{1}{\sqrt{3}}\, \frac{f_{\hpi}}{\mathrm{GeV}} \bigg[\frac{\kappa_1^2}{(M_{X_1}/\mathrm{TeV})^2} -\frac{\kappa_3^2}{(M_{X_3}/\mathrm{TeV})^2} \bigg]  .
\end{equation}
%\frac{1}{\mathrm{PeV}}   \bigg(\frac{f_{\hpi}}{\mathrm{GeV}} \bigg)
In Eq.~\eqref{eq:single_J_terms}, the $\mathcal{O}(J_\mu \hpi \partial^\mu \hpi)$ terms, which would mediate in particular DM-nucleon scattering, vanish as a consequence of the $\mathcal{C}$ symmetry. In summary, we find the effective couplings of the unstable $\he$ to SM fermions,
\begin{equation}
\mathcal{L}_{\he} = -\, \frac{\partial_\mu \he_0}{F_{\he}} \sum_{f\,\in\, q,\, l} c_f  \bar{f} \gamma^\mu \gamma_5 f\,, \qquad c_d = -\,1/2 \,,\quad c_f= 0\;\; \mathrm{if}\; f \neq d\,.
\end{equation}
The corresponding pattern of $\he$ decays to SM hadrons and $\gamma\gamma$ can be evaluated with the results of Ref.~\cite{Balkin:2025enj}.

At the LHC, $\psi\overline{\psi}$ pairs are produced via $t$-channel $X$ exchange, while $XX^\ast$ pair production is mediated by SM QCD interactions and by $t$-channel $\psi$ exchange. In addition, $X\psi$ can be produced via an $s$-channel SM quark. Once produced, the heavy bi-fundamental scalars $X$ decay to one SM quark and one dark quark, which turn into a SM jet and a dark jet, respectively. If dark jets are semi-visible, a recent LHC search covers the range $r_{\rm inv}\in[0.1,0.9]$ for $\psi\overline{\psi}$ production with a sensitivity of $\mathcal{O}(10)$~fb, depending on the $X$ mass~\cite{ATLAS:2023swa}. 
Several searches were conducted for final states containing both emerging jets and SM jets, resulting from $XX^\ast$ QCD pair production, assuming macroscopic decay lengths for the dark mesons~\cite{CMS:2018bvr,CMS:2024gxp,ATLAS:2025lfx}. The optimal upper limit reaches the $\lesssim \mathcal{O}(0.1)$~fb level, depending on the dark meson lifetime, which would correspond to a constraint on $M_X$ close to 2 TeV. However, similarly to the $Z'$-portal scenario discussed in Sec.~\ref{sec:zprime}, these searches do not consider large fractions of invisible dark mesons, and are therefore expected to lose sensitivity when applied to the present model, where $r_{\rm inv} \gtrsim  0.8$. A search re-optimization may be needed in this case. For even longer lifetimes, the dark jets are fully invisible and the signal becomes jets + missing transverse energy (MET), analogous to squark pair production~\cite{CMS:2019zmd,ATLAS:2020syg}. The current bound on $M_X$ would be at least 1.5~TeV in this regime.

One can also consider couplings to other SM fermions in Eq.~\eqref{eq:bifundamental}. If the dark quarks couple only to one heavy SM quark, $b$ or $t$, the leading interaction mediating $\he$ decays will be $\he_0 G_{\mu\nu}\widetilde{G}^{\mu\nu}$, generated by integrating out the heavy quark at one loop. The corresponding pattern of decays to SM hadrons was presented in Ref.~\cite{Balkin:2025enj}. One could also consider couplings to SM leptons. In that case, the $X$ scalars are produced through electroweak interactions at the LHC, with smaller cross sections. On the other hand, $\he$ decays predominantly to lepton pairs, potentially leading to displaced lepton-jet signals~\cite{ATLAS:2014fzk,Diamond:2017ohe}.

\section{Conclusions}
\label{sec:conclusions}

In this work, we entertained the possibility that DM emerges from a confining dark sector, neutral under the SM gauge interactions, in the form of pNGB mesons (dark pions). We presented {\it slide dark matter}, a generic thermal mechanism that allows dark pions with GeV-scale masses to acquire the observed relic density. The stability of the dark pions is ensured by a flavor symmetry (analogous to the isospin symmetry of SM QCD, but exact). Heavier pNGB dark mesons are also present in the spectrum and are either stable or unstable, depending on whether they are charged or neutral under the flavor symmetry. The DM relic density is determined by the interplay of up-scatterings of the dark pions to heavier dark mesons, and decays of the unstable dark mesons to SM particles.

The slide DM mechanism typically requires at least three types of dark mesons: $\hpi$, $\hK$, and $\he$, in analogy to the $\pi$, $K$, and  $\eta$ of the SM sector. As an illustration, in this work we focused on a minimal model based on the $SO(N_d)$ dark gauge group and the $SU(3)/SO(3)$ coset, though the mechanism can be generalized to other gauge groups and symmetry breaking patterns. DM mainly consists of the lightest $\hpi$ fields, but $\hK$ plays an important role in converting between $\hpi$ and the unstable $\he$. We found that for large $\he$ decay widths, the DM relic density is determined by the freeze-out of the $\hpi\hpi \to \hK\hK$ up-scattering process. By contrast, for small $\he$ decay widths, the relic density is controlled by the decoupling of $\he$ decays to SM particles. The transition typically occurs in the range of $\Gamma_{\he}^{-1} \sim 10\,$-$\,10^3\;\mathrm{cm}$, depending on other model parameters. The (inverse) decay of $\he$ to SM particles also keeps the dark and SM sectors in kinetic equilibrium before freeze-out or decoupling, hence the decay width cannot be too small, with $\Gamma_{\he}^{-1} \lesssim 10^3\;\mathrm{m}$. This is a favourable outcome for searches of $\he$ decays with accelerator experiments.

One important feature of the model is that a dark charge conjugation symmetry forbids any vector current interactions between the dark mesons and the SM fields. As a result, the signals in both direct and indirect searches for DM are strongly suppressed in most of the parameter space. Only when the $\hK$'s comprise a significant fraction of the DM density, which occurs if the mass splittings among the dark mesons are small ($\ll 10\%$), does the $\hK\hK\to \hpi \he$ annihilation followed by $\he$ decay to SM particles yield potentially detectable signals in indirect searches for DM. In the rest of the parameter space, particle accelerators typically offer the only viable path to discovery, with the (HL)-LHC playing the main role in the near- and mid-term future.

At the LHC, the expected signatures depend strongly on the UV completion of the portal interaction that mediates $\he$ decays to SM particles. In a given UV completion, dark quarks are generally produced with energies much larger than the dark confinement scale, resulting in dark showers made (mostly) of dark mesons. It is expected that $\lesssim 1/5$ of the produced dark mesons are unstable $\he$'s, while the rest are invisible. If $\he$ is light enough, it becomes an LLP, leading to semi-visible emerging jet signals. In this work we showed that three well-known classes of mediators can UV-complete slide DM, and outlined the corresponding signatures. If the $Z$ boson acts as the mediator, its decays produce final states with relatively low dark meson multiplicity and soft momenta. Specialized search strategies are then required to retain sensitivity to the (dimuon) displaced vertices generated by $\he$ decays. For heavy mediators, namely a $Z'$ vector boson or scalar fields charged under both dark color and SM color, modification of current search strategies is needed to identify the semi-visible emerging jets. For $\he$ decay lengths much longer than $1$~meter, the proposed LHC auxiliary detectors for LLP searches will provide complementary coverage.

The slide DM mechanism provides a compelling link between GeV-scale thermal DM and dark shower signatures at the LHC. It can serve as a benchmark to compare the reach of different experimental searches, and its phenomenology warrants more detailed studies. Finally, a wider spectrum of accelerator experiments~\cite{BESIII:2009fln,SHiP:2015vad,NA62:2017rwk,Coloma:2023oxx,Niedziela:2024khw,Ai:2025xop,Ai:2025cpj,FCC:2025lpp} may extend the sensitivity to the mechanism even further, a topic which is left for future investigation.

\section*{Acknowledgments}
We thank Haolin Li, Bingxuan Liu, Riccardo Rattazzi and Xingbo Yuan for useful discussions.
HC is supported by the US Department of Energy grant No.~DE-SC0009999. HC's work was performed in part at the Munich Institute for Astro-, Particle and BioPhysics (MIAPbP) which is funded by the Deutsche Forschungsgemeinschaft (DFG, German Research Foundation) under Germany´s Excellence Strategy – EXC-2094 – 390783311, Aspen Center for Physics, which is supported by National Science Foundation grant PHY-2210452, and Institute of Physics, Academia Sinica. ES is supported by the grant RYC2023-042775-I, funded by the Spanish Ministry of Science, Innovation and Universities (MCIU) through the Spanish State Research Agency (AEI, 10.13039/501100011033) and by the FSE+. The work of ES was also performed in part at the Aspen Center for Physics.

\appendix
\section{Boltzmann Equations}
\label{app:Boltzmann}
In this appendix we present the complete Boltzmann equations that govern the thermal evolution of the dark mesons in the $SU(3)/SO(3)$ model. 
\subsection{Thermally-averaged cross sections for $2\to 2$ and $3 \to 2$ processes}
The leading interactions involving four dark pNGBs are obtained from the chiral Lagrangian in Eq.~\eqref{eq:p2},
\begin{align}
&\mathcal{L}_{\hpi, \,p^2} \supset \frac{1}{24 f_{\hpi}^2} \bigg\{ m^2_{\hpi}  \Big[\he_0^2 + 2 \hpi_+ \hpi_- + 2 \hK_{+1/2} \hK_{-1/2} \Big]^2 \hspace{-1.5mm} + 2\big(m_{\hK}^2 - m_{\hpi}^2 \big)  \Big[ \frac{8}{9} \he_0^4 + 3 \he_0^2 \hK_{+1/2} \hK_{-1/2} \nonumber \\ 
-\;& \sqrt{\frac{2}{3}} \big( \he_0 \hK_{+1/2}^2 \hpi_- + \he_0 \hK_{-1/2}^2 \hpi_+ \big) + 2 \hK_{+1/2}^2 \hK_{-1/2}^2 + 2 \hK_{+1/2} \hK_{-1/2} \hpi_+ \hpi_- \Big]  \bigg\} \nonumber \\
+ \;& \frac{1}{24 f_{\hpi}^2}  \bigg\{ 4 (\partial \hpi_{+})^2 \hpi_-^2 + 4 \hpi_+^2 (\partial \hpi_-)^2  - 8 \partial \hpi_+ \hpi_+ \partial \hpi_-  \hpi_- + (\partial \hK_{+ 1/2})^2 \hK_{- 1/2}^2  + \hK_{+ 1/2}^2 (\partial \hK_{- 1/2})^2 
\nonumber \\ 
- \;& 2 \partial \hK_{+ 1/2} \hK_{+ 1/2} \partial \hK_{- 1/2} \hK_{- 1/2} 
+  8 \hK_{+1/2} \partial \hK_{-1/2}  \hpi_+ \partial \hpi_- + 8 \partial \hK_{+1/2} \hK_{-1/2}  \partial \hpi_+ \hpi_- \nonumber \\
- \;& 4 \hK_{+1/2} \hK_{-1/2}   \partial \hpi_+ \partial \hpi_-  -4 \hK_{+1/2} \partial \hK_{-1/2}   \partial \hpi_+ \hpi_- 
-4 \partial \hK_{+1/2}  \hK_{-1/2} \hpi_+ \partial \hpi_- 
\nonumber \\ 
- \;& 4 \partial \hK_{+1/2} \partial \hK_{-1/2}   \hpi_+ \hpi_- 
-\hspace{-0.3mm} 6 \he_0^2 \partial \hK_{+1/2} \partial \hK_{-1/2}   - \hspace{-0.3mm} 6 (\partial \he_0)^2  \hK_{+1/2} \hK_{-1/2} + \hspace{-0.3mm} 6 \he_0 \partial \he_0 \partial \hK_{+1/2}  \hK_{-1/2} \nonumber \\ 
+ \;& 6 \he_0 \partial \he_0 \hK_{+1/2} \partial \hK_{-1/2} 
+ 2 \sqrt{6} \Big[ \partial \he_0 \partial \hK_{-1/2} \hK_{-1/2} \hpi_+ + \partial \he_0 \partial \hK_{+1/2} \hK_{+1/2} \hpi_- - \partial \he_0  \hK_{+1/2}^2 \partial \hpi_- \nonumber \\ 
-\;& \partial \he_0  \hK_{-1/2}^2 \partial \hpi_+  - \he_0 (\partial \hK_{-1/2})^2  \hpi_+ - \he_0 (\partial \hK_{+1/2})^2  \hpi_- + \he_0 \partial \hK_{-1/2}   \hK_{-1/2}  \partial \hpi_+ \nonumber \\
&\hspace{9cm} + \he_0 \partial \hK_{+1/2} \hK_{+1/2} \partial \hpi_- \Big] \bigg\}\,.
\label{eq:L4point}
\end{align}
The above interactions mediate $2\to 2$ scatterings among the pNGBs. We focus on the low-energy regime, well below the scalar and vector resonances, where Eq.~\eqref{eq:L4point} provides a good description. All the $2\to 2$ scattering processes proceed through the leading $s$-wave, and subleading contributions are neglected.

We start with $\he_0 \he_0 \to \hpi_+ \hpi_{-}\,$, mediated only by the mass terms in $\mathcal{L}_{\hpi, \,p^2}$, finding for the thermally-averaged cross section
\begin{equation}
\langle \sigma v \rangle_{\he \he \to \hpi \hpi} \simeq \frac{m_{\hpi}^4}{288\pi f_{\hpi}^4 m_{\he}^2 } \bigg( 1 - \frac{m_{\hpi}^2}{m_{\he}^2}\bigg)^{1/2}~,
\end{equation}
where $v$ is the relative velocity. For $\hK_{\pm 1/2}\he_0 \to \hK_{\mp 1/2} \hpi_\pm$, which is mediated both by the mass terms and derivative terms in $\mathcal{L}_{\hpi, \,p^2}$, we have
\begin{align}
\langle \sigma v \rangle_{\hK_{ }\he \to \hK_{ } \hpi_{}} \simeq\;& \frac{ \Big[ 3 m_{\he}^3+12 m_{\he}^2 m_{\hK} + 8 m_{\he} m_{\hK}^2 + m_{\hpi}^2 \big(m_{\he} - 8 m_{\hK}\big) + 8 m_{\hK}^3\Big]^2}{6912 \pi  f_{\hpi}^4 m_{\he} m_{\hK} (m_{\he} + m_{\hK})^4} \nonumber\\ 
&\hspace{4cm}\times  \Big[\big(m_{\he}^2 - m_{\hpi}^2\big) \big( \big(m_{\he} + 2 m_{\hK}\big)^2 - m_{\hpi}^2\big) \Big]^{1/2}~,
\end{align}
while for $\hK_{\pm 1/2} \hK_{\pm 1/2}  \to  \hpi_\pm \he_0$ we obtain
\begin{equation}
\langle \sigma v \rangle_{\hK_{} \hK_{} \to  \hpi_{} \he} \simeq \frac{\big(28 m_{\hK}^2 - 3 m_{\he}^2 - m_{\hpi}^2 \big)^2}{27648 \pi  f_{\hpi}^4 m_{\hK}^4} \Big[ m_{\he}^4 - 2 m_{\he}^2 \big(4 m_{\hK}^2 + m_{\hpi}^2 \big) + \big( 4 m_{\hK}^2 - m_{\hpi}^2 \big)^2 \Big]^{1/2} \,.
\end{equation}
For $\hK_{+1/2} \hK_{-1/2} \to \hpi_+ \hpi_-$ we find
\begin{equation}
\langle \sigma v \rangle_{\hK_{ } \hK_{ } \to \hpi \hpi} \simeq \frac{m_{\hK}^2 }{32 \pi  f_{\hpi}^4} \bigg(1 - \frac{m_{\hpi}^2}{m_{\hK}^2} \bigg)^{1/2}~,
\end{equation}
and finally for $\he_0 \he_0 \to \hK_{+1/2} \hK_{-1/2}\,$,
\begin{equation}
\langle \sigma v \rangle_{\he \he \to \hK_{ } \hK_{ }} \simeq \frac{  \big(   15 m_{\he}^2 - m_{\hpi}^2 \big)^2}{1152 \pi  f_{\hpi}^4 m_{\he}^2} \bigg( 1 - \frac{m_{\hK}^2}{m_{\he}^2} \bigg)^{1/2}\,.
\end{equation}
The $3 \to 2$ processes arise from the WZW term in Eq.~\eqref{eq:WZW}, with $d$-wave thermally-averaged cross sections given by  
\begin{equation}\label{eq:3_2}
\langle \sigma v^2 \rangle_{123\to 45} \simeq \frac{N_d^2 T^2}{6144\pi^5 f_{\hpi}^{10}}\, \frac{\lambda^{3/2}[(m_1+m_2+m_3)^2, m_4^2, m_5^2]}{ (m_1+m_2+m_3)^3}\;,
\end{equation}
where
\begin{equation}\label{eq:lambda_def}
\lambda[a,b,c] \equiv a^2 + b^2 + c^2 - 2 ab - 2 ac - 2 bc\,.
\end{equation}

\subsection{Complete Boltzmann equations}
\label{app:Boltzmann_eqs}
Using $x = m_{\hpi}/T$ as time variable, the Boltzmann equations for $Y_{\hpi}$, $Y_{\hK}$, and $Y_{\he}$ are given by
\begin{align}
&\frac{dY_{\hpi}}{dx} = -\, \frac{s(x)}{\tilde{H}(x) x} \Bigg[ - \langle \sigma v \rangle_{\hK \hK \to \hpi \hpi} \bigg( Y_{\hK}^2 - Y^{\rm eq\,2}_{\hK} \frac{Y_{\hpi}^2}{Y_{\hpi}^{\rm eq\,2}}  \bigg) - \frac{1}{2} \langle \sigma v \rangle_{\hK \hK \to  \hpi \he} \bigg( Y_{\hK}^2 - Y^{\rm eq\;2}_{\hK} \frac{Y_{\hpi}Y_{\he}}{Y_{\hpi}^{\rm eq}Y_{\he}^{\rm eq}}  \bigg)  \nonumber \\
&- \langle \sigma v \rangle_{\hK \he \to \hK \hpi} \bigg( Y_{\hK} Y_{\he} -  Y_{\he}^{\rm eq} \frac{Y_{\hK}Y_{\hpi}}{Y_{\hpi}^{\rm eq}}  \bigg) - \frac{1}{2} \langle \sigma v \rangle_{ \he  \he \to \hpi \hpi} \bigg( Y_{\he}^2  - Y_{\he}^{\rm eq\;2} \frac{ Y_{\hpi}^2 }{Y_{\hpi}^{\rm eq\,2} }  \bigg) \Bigg] \nonumber
\end{align}
\begin{align}
-&\frac{s(x)^2}{\tilde{H}(x) x} \Bigg[ 2\,  \langle \sigma v^2 \rangle_{\hK \hpi \hpi \to \he \hK} \bigg( Y_{\hK} Y_{\hpi}^2 - Y_{\hpi}^{\rm eq\,2} \frac{Y_{\he} Y_{\hK}}{Y_{\he}^{\rm eq} } \bigg) + \langle \sigma v^2 \rangle_{\he \hpi \hpi \to  \hK \hK} \bigg( Y_{\he} Y_{\hpi}^2 - Y_{\he}^{\rm eq} Y_{\hpi}^{\rm eq\,2} \frac{Y_{\hK}^2}{Y_{\hK}^{\rm eq\,2} } \bigg)\nonumber \\ 
&\hspace{6cm}- \langle \sigma v^2 \rangle_{\he \hK \hK \to \hpi \hpi} \bigg( Y_{\he} Y_{\hK}^2 - Y_{\he}^{\rm eq} Y_{\hK}^{\rm eq\,2} \frac{Y_{\hpi}^2}{Y_{\hpi}^{\rm eq\,2} } \bigg) \Bigg]  ,  \label{eq:Boltz_pi}
\end{align}
\begin{align}
&\,\frac{dY_{\hK}}{dx} = -\, \frac{s(x)}{\tilde{H}(x) x} \Bigg[ \langle \sigma v \rangle_{\hK \hK \to \hpi \hpi} \bigg( Y_{\hK}^2 - Y^{\rm eq\,2}_{\hK} \frac{Y_{\hpi}^2}{Y_{\hpi}^{\rm eq\,2}}  \bigg) + \langle \sigma v \rangle_{\hK \hK \to  \hpi \he}  \bigg( Y_{\hK}^2 - Y^{\rm eq\;2}_{\hK} \frac{Y_{\hpi}Y_{\he}}{Y_{\hpi}^{\rm eq}Y_{\he}^{\rm eq}}  \bigg)\hspace{4mm} \nonumber  \\ 
&\hspace{-3mm}-  \frac{1}{2} \langle \sigma v \rangle_{ \he  \he \to \hK \hK} \bigg( Y_{\he}^2  - Y_{\he}^{\rm eq\;2} \frac{ Y_{\hK}^2 }{Y_{\hK}^{\rm eq\,2} }   \bigg) \Bigg] - \frac{s(x)^2}{\tilde{H}(x) x} \Bigg[ 2\, \langle \sigma v^2 \rangle_{\hK \hK \hpi \to \he \hpi} \bigg( Y_{\hK}^2 Y_{\hpi} - Y_{\hK}^{\rm eq\,2} \frac{Y_{\he} Y_{\hpi}}{Y_{\he}^{\rm eq} } \bigg) \nonumber \\
&\hspace{-3mm}- \hspace{-0.5mm} \langle \sigma v^2 \rangle_{\he \hpi \hpi \to  \hK \hK} \bigg( \hspace{-0.5mm}Y_{\he} Y_{\hpi}^2 \hspace{-0.5mm}-\hspace{-0.5mm} Y_{\he}^{\rm eq} Y_{\hpi}^{\rm eq\,2} \frac{Y_{\hK}^2}{Y_{\hK}^{\rm eq\,2} } \hspace{-0.5mm}\bigg) \hspace{-0.5mm}+\hspace{-0.5mm} \langle \sigma v^2 \rangle_{\he \hK \hK \to \hpi \hpi} \bigg(\hspace{-0.5mm} Y_{\he} Y_{\hK}^2 -\hspace{-0.5mm} Y_{\he}^{\rm eq} Y_{\hK}^{\rm eq\,2} \frac{Y_{\hpi}^2}{Y_{\hpi}^{\rm eq\,2} }\hspace{-0.5mm} \bigg)\hspace{-0.25mm}\Bigg], \label{eq:Boltz_K} 
\end{align}
\begin{align}
&\frac{dY_{\he}}{dx} = -\, \frac{s(x)}{\tilde{H}(x) x} \Bigg[\hspace{-0.5mm} -\hspace{-0.5mm} \langle \sigma v \rangle_{\hK \hK \to  \hpi\he} \bigg(\hspace{-0.5mm} Y_{\hK}^2 - Y^{\rm eq\;2}_{\hK} \frac{Y_{\hpi}Y_{\he}}{Y_{\hpi}^{\rm eq}Y_{\he}^{\rm eq}} \hspace{-0.5mm} \bigg) + 2\, \langle \sigma v \rangle_{\hK \he \to \hK \hpi} \bigg(\hspace{-0.5mm} Y_{\hK} Y_{\he} - Y_{\he}^{\rm eq} \frac{Y_{\hK}Y_{\hpi}}{Y_{\hpi}^{\rm eq}}\hspace{-0.5mm}\bigg) \nonumber   \\ 
&+  \langle \sigma v \rangle_{ \he  \he \to \hK \hK} \bigg( Y_{\he}^2  - Y_{\he}^{\rm eq\;2} \frac{ Y_{\hK}^2 }{Y_{\hK}^{\rm eq\,2} }  \bigg) + \langle \sigma v \rangle_{ \he  \he \to \hpi \hpi} \bigg( Y_{\he}^2  - Y_{\he}^{\rm eq\;2} \frac{ Y_{\hpi}^2 }{Y_{\hpi}^{\rm eq\,2} }  \bigg) \Bigg]
- \frac{\langle \Gamma \rangle_{\he}}{\tilde{H}(x)x}  \big(Y_{\he} - Y_{\he}^{\rm eq} \big) \nonumber \\
&- \frac{s(x)^2}{\tilde{H}(x) x} \Bigg[ \hspace{-0.5mm}-\hspace{-0.5mm} 2\, \langle \sigma v^2 \rangle_{\hK \hpi \hpi \to \he \hK} \bigg(\hspace{-0.5mm} Y_{\hK} Y_{\hpi}^2 - Y_{\hpi}^{\rm eq\,2} \frac{Y_{\he} Y_{\hK}}{Y_{\he}^{\rm eq} } \hspace{-0.5mm}\bigg) - 2\, \langle \sigma v^2 \rangle_{\hK \hK \hpi \to \he \hpi} \bigg( Y_{\hK}^2 Y_{\hpi} - Y_{\hK}^{\rm eq\,2} \frac{Y_{\he} Y_{\hpi}}{Y_{\he}^{\rm eq} } \bigg)  \nonumber \\
&+ \langle \sigma v^2 \rangle_{\he \hpi \hpi \to  \hK \hK} \bigg( Y_{\he} Y_{\hpi}^2 - Y_{\he}^{\rm eq} Y_{\hpi}^{\rm eq\,2} \frac{Y_{\hK}^2}{Y_{\hK}^{\rm eq\,2} } \bigg)  + 4\, \langle \sigma v^2 \rangle_{\he \hK \hpi \to \hK \hpi} \Big( Y_{\he} Y_{\hK} Y_{\hpi} - Y_{\he}^{\rm eq} Y_{\hK} Y_{\hpi} \Big) \nonumber \\  
&\hspace{6cm}+ \langle \sigma v^2 \rangle_{\he \hK \hK \to \hpi \hpi} \bigg( Y_{\he} Y_{\hK}^2 - Y_{\he}^{\rm eq} Y_{\hK}^{\rm eq\,2} \frac{Y_{\hpi}^2}{Y_{\hpi}^{\rm eq\,2} } \bigg) \Bigg]\,, \label{eq:Boltz_eta}
\end{align}
where
\begin{align}
 s(x) &= \frac{2\pi^2}{45} g_{\ast s} (x) \frac{m_{\hpi}^3}{x^3}\,,\qquad \tilde{H}(x) = \frac{H(x)}{1 - \frac{1}{3} \frac{d\log g_{\ast s}}{d \log x}} \,, \qquad H(x) = \frac{\pi g_\ast (x)^{1/2} m_{\hpi}^2}{3\sqrt{10}\hspace{0.3mm} M_{\rm Pl} x^2}\;, \nonumber \\ 
&Y^{\rm eq}_X (x) \simeq \frac{1}{s(x)}\frac{m_X^2 m_{\hpi}}{2\pi^2 x}\, K_2\hspace{-0.5mm} \left( \frac{m_X }{m_{\hpi}}x\right)\,, \qquad ~\langle \Gamma \rangle_{\he} \simeq  \frac{K_1 \big( (1+\Delta_{\he})x \big)}{K_2 \big((1+\Delta_{\he})x\big)}\,\Gamma_{\he}~,
\end{align}
with $K_{1,2}$ denoting modified Bessel functions of the second kind. In Eqs.~\eqref{eq:Boltz_pi},~\eqref{eq:Boltz_K} and~\eqref{eq:Boltz_eta} the $U(1)_{T_2}$ charges were suppressed, exploiting the fact that the number densities of conjugate states and the cross sections of conjugate processes are the same, \ie~$Y_{\hpi} = Y_{\hpi_+}=Y_{\hpi_-}$and $\langle \sigma v \rangle_{\hK \he \to \hK \hpi} = \langle \sigma v \rangle_{\hK_{+1/2} \he_0 \to \hK_{-1/2} \hpi_+}= \langle \sigma v \rangle_{\hK_{-1/2} \he_0 \to \hK_{+1/2} \hpi_-}$, and so on.

At low temperatures, after the $3\to 2$ processes have frozen out and can therefore be ignored, it is easy to check that the total comoving number density of the dark mesons obeys 
\begin{equation}
\frac{d}{dx} Y_{\rm tot} = \frac{d}{dx}\big(2 Y_{\hpi} + 2 Y_{\hK} + Y_{\he} \big) = -\, \frac{\langle \Gamma \rangle_{\he}} {\tilde{H}(x)x}\big(Y_{\he}-Y_{\he}^{\rm eq}\big)  \simeq -\,\frac{ \Gamma_{\he} } {\tilde{H}(x)x} \big(Y_{\he}-Y_{\he}^{\rm eq}\big) \,, 
\end{equation}
which is Eq.~\eqref{eq:Gammaeff2}, reflecting the fact that only $\he$ decays can change the total number of the dark mesons.

\section{Higgs Interactions in the $Z$-Portal Model}
\label{app:Higgs}
Here we discuss the effects of the operators involving the Higgs field that appear in the $Z$-portal model, see the last two lines of Eq.~\eqref{eq:EFT_total}. We begin by considering the operators in the second line alone, which dominate if the Yukawa couplings $y_i$ and $\tilde{y}_i$ have comparable sizes. At the end we comment on the operators in the third line, which are extra suppressed by the ratio of the light and heavy dark quark masses $m_i/M_{Q_i}$, but survive even in the hierarchical limit, \eg~$\tilde{y}_i \to 0$.

Focusing on the operators in the second line of Eq.~\eqref{eq:EFT_total}, the total dark quark masses need to satisfy $|m_i - y_i \tilde{y}_i v^2/M_{Q_i}| \ll \widehat{\Lambda}$, where $\widehat{\Lambda}$ is the dark QCD confinement scale, to ensure the presence of light pNGB mesons in the dark hadron spectrum. Additionally, the same operators induce decays of the Higgs boson to dark quarks, which produce dark showers. The corresponding partial width,
\begin{equation}
    \Gamma(h \to \bm{\psi} \overline{\bm{\psi}}\hspace{0.2mm}) = m_h  \frac{N_d  v^2 }{2\pi} \bigg( \frac{|y_1 \tilde{y}_1|^2}{M_{Q_1}^2} + \frac{|y_3 \tilde{y}_3|^2}{2 M_{Q_3}^2} \bigg)\,,
\end{equation}
should not exceed the upper bound on the Higgs branching ratio to undetected final states from LHC Run 2 measurements, which is $0.21$ at $95\%$~CL~\cite{CMS:2026nce}. This yields the constraint
\begin{equation}\label{eq:h_decay_undet}
  \bigg( \frac{|y_1 \tilde{y}_1|^2}{M_{Q_1}^2} + \frac{|y_3 \tilde{y}_3|^2}{2 M_{Q_3}^2} \bigg)^{1/2} < 0.012 \;\mathrm{TeV}^{-1} \bigg( \frac{5}{N_d} \bigg)^{1/2}\,.
\end{equation}
The displaced decays of $\he$'s produced in Higgs-initiated dark showers have also been probed by the LHC experiments~\cite{Cheng:2024aco}. The CMS search for displaced dimuon resonances based on data scouting~\cite{CMS:2021sch} reached sensitivities $\sim 10^{-4}$ on $\mathrm{BR}_h \big(\bm{\psi} \overline{\bm{\psi}}\hspace{0.2mm}\big)\times \mathrm{BR}_{\he} (\mu^+\mu^-)$, for proper decay lengths of $\he$ in the optimal mm$\,$-$\,$cm range. Auxiliary detectors located far away from the LHC interaction points can be sensitive to longer decay lengths. 

Upon confinement, the $h\psi_i \psi_i$ couplings turn into Higgs-dark meson couplings. This is captured by the following replacement in the chiral Lagrangian, Eq.~\eqref{eq:p2},
\begin{equation}
\mathcal{M} \to \mathcal{M} - (v + h)^2 \,\mathrm{diag}\bigg( \frac{y_1 \tilde{y}_1}{M_{Q_1}}, \frac{y_1 \tilde{y}_1}{M_{Q_1}}, \frac{y_3 \tilde{y}_3}{M_{Q_3}} \bigg)\,.
\end{equation}
For the dominant component of DM, $\hpi_{\pm}$, we obtain the interactions
\begin{equation}\label{eq:h_DM_DM}
\mathcal{L}_{\hpi,\,p^2} \supset B\, \frac{y_1 \tilde{y}_1 + y_1^\ast \tilde{y}_1^\ast}{M_{Q_1}}\,  \hpi_+ \hpi_- (v+h)^2\,,
\end{equation}
which include an $h \hpi_+ \hpi_-$ coupling. Hence, the tree-level Higgs exchange leads to spin-independent (SI) DM scattering on nucleons, with cross section
\begin{equation}
\sigma_{\rm SI}^{\hpi_\pm N} \simeq \frac{1}{4\pi} \bigg( \frac{m_{\hpi} m_N}{m_{\hpi} + m_N} \bigg)^2 \frac{m_N^2}{m_{\hpi}^2}  \frac{f_N^2}{m_h^4} \bigg( 2 B\, \frac{y_1 \tilde{y}_1 + y_1^\ast \tilde{y}_1^\ast}{M_{Q_1}} \bigg)^2\,, \quad f_N = \frac{2}{9} + \frac{7}{9}\sum_{q \,=\, u, d, s} f_{T_q}\,,
\end{equation}
where $m_N$ is the average nucleon mass and the values of the $f_{T_q}$ can be found in Ref.~\cite{FlavourLatticeAveragingGroupFLAG:2024oxs}, resulting in $f_N \approx 0.31$. We then arrive at
\begin{equation}\label{eq:SI_pi}
\sigma_{\rm SI}^{\hpi_\pm N} \simeq 6.2\times 10^{-46}\;\mathrm{cm}^2\, b^2 \bigg( \frac{1.94}{m_{\hpi}/\mathrm{GeV} + 0.94} \bigg)^2 \bigg( \frac{f_{\hpi}}{\mathrm{GeV}}\bigg)^2 \bigg( \frac{y_1 \tilde{y}_1 + y_1^\ast \tilde{y}_1^\ast}{0.02}\bigg)^2 \bigg( \frac{\mathrm{TeV}}{M_{Q_1}}\bigg)^2\,,
\end{equation}
where $b \equiv B/(4 \pi f_{\hpi})$ is expected to be an $\mathcal{O}(1)$ number. For the ratio $(y_1 \tilde{y}_1 + y_1^\ast \tilde{y}_1^\ast)/M_{Q_1}$, we take a reference value close to the upper bound from Higgs undetected decays in Eq.~\eqref{eq:h_decay_undet}. The reference scattering cross section in Eq.~\eqref{eq:SI_pi} is within the neutrino fog~\cite{OHare:2021utq} for DM masses $m_{\hpi} \lesssim 8\;\mathrm{GeV}$, but some experimental sensitivity is already present if $f_{\hpi}$ is moderately larger than $1$~GeV (see, for instance, the most recent results reported by LZ~\cite{LZ:2025igz}). Thus, if the dark sector couples to the SM via the $Z$ portal, a signal in direct DM searches is possible. Yet, this signal can be absent if $Q_{1,2}^{(c)}$ are decoupled, or if the Yukawas are strongly hierarchical (for example, $\tilde{y}_i \to 0$).

Finally, we discuss the operators in the third line of Eq.~\eqref{eq:EFT_total}, which survive even for strongly hierarchical Yukawas. In this case, the upper bound on the Higgs branching ratio to undetected final states only imposes a very loose requirement on the size of the underlying parameters, as Eq.~\eqref{eq:h_decay_undet} is replaced by
\begin{equation}\label{eq:h_decay_undet_subleading}
  \Bigg[ \frac{(|y_1|^2 + |\tilde{y}_1|^2)^2 m_1^2}{M_{Q_1}^4} + \frac{(|y_3|^2 + |\tilde{y}_3|^2)^2 m_3^2}{2 M_{Q_3}^4} \Bigg]^{1/2} < 0.024 \;\mathrm{TeV}^{-1} \bigg( \frac{5}{N_d} \bigg)^{1/2}\,.
\end{equation}
The corresponding SI cross section for DM scattering on nucleons,
\begin{equation}\label{eq:SI_pi_subleading}
\sigma_{\rm SI}^{\hpi_\pm N} \simeq 2.5\times 10^{-51}\;\mathrm{cm}^2\, \bigg( \frac{1.94}{m_{\hpi}/\mathrm{GeV} + 0.94} \bigg)^2 \bigg( \frac{m_{\hpi}}{\mathrm{GeV}}\bigg)^4 \big( |y_1|^2 + |\tilde{y}_1|^2 \big)^2 \bigg( \frac{\mathrm{TeV}}{M_{Q_1}}\bigg)^4\,,
\end{equation}
is well inside the neutrino fog and seemingly out of near-term experimental reach. 

\section{Further Description of the LHC Sensitivity Projections}
\label{app:phenomenology}

To estimate the projected LHC sensitivity to dark shower signals in the $Z$-portal scenario, see Sec.~\ref{ssec:Zportal}, we follow the results of Ref.~\cite{Cheng:2024aco} but with updated $\he$ decay branching ratios~\cite{Balkin:2025enj}. In particular, the projected CMS limits were recast from the data scouting search for LLPs decaying into muon pairs in the tracker~\cite{CMS:2021sch}, making some conservative assumptions. Very recently, the CMS collaboration has published a dark shower search for low-mass dimuon DVs using the data parking technique~\cite{CMS:2025fnr}. The technique separates the steps of raw data storage and processing, effectively reducing the muon threshold, which allows CMS to probe the low-$m_{\mu\mu}$ region. The search was based on a Higgs portal model. A direct comparison between the Higgs portal results in Refs.~\cite{CMS:2025fnr} and \cite{Cheng:2024aco} indicates that the data parking sensitivity could be $3\,$-$\,4$ times better than our conservative projections based on the data scouting search. However, since machine learning techniques were used in key steps of the data parking analysis, recasting the bounds of Ref.~\cite{CMS:2025fnr} to the $Z$-portal scenario is not straightforward. We leave this for future work.

\begin{table}[t]
\centering
\begin{footnotesize}
\begin{tabular}{cccccc}
%\hline
Detector & Geometry  & Displacement (m) & Dimensions (m) & Luminosity (fb$^{-1}$)\\
\hline
FASER 2~\cite{FASER:2018eoc} & cylinder & 0, 0, 480 & 2, 5 & 3000\\
MATHUSLA~\cite{MATHUSLA:2025zyt}  & box & 88.5, 0, 90 & 11, 40, 40 & 3000\\
Codex-b~\cite{CODEX-b:2025rck} & box & 31, 2, 10 & 10, 10, 10 & 300\\
ANUBIS (ceiling)~\cite{ANUBIS:2025sgg} & special & $1.7$, $0$, $\sim 19.5$  & $\sim 14,~\sim 21,~\sim 53 $ & 3000 \\
ANUBIS (shaft)~\cite{ANUBIS:2025sgg} & cylinder & 1.7, 51.5, 13.25 & 17.5, 57 & 3000 \\
\hline
\end{tabular}
\end{footnotesize}
\caption{Simplified description of the LHC auxiliary detectors considered in this work. The third column shows the approximate distance of the geometric center of the detector from the closest LHC interaction point in 3 dimensions, with the last number always corresponding to the beam direction. The detector dimensions are also listed, in the fourth column. 
For cylindrical detectors, we quote the diameter and length instead. The last column lists the expected integrated luminosity in the HL-LHC era. The MATHUSLA geometry corresponds to the updated scaled-back design~\cite{MATHUSLA:2025zyt}. The Codex-b parameters are taken from the original design, as a new design is not available yet. For ANUBIS, we take the ceiling configuration as our benchmark. The effective volume geometry of this design is asymmetric and cannot be simply described by a few numbers; in this case, we list approximate values for the displacement of the geometric center and the dimensions. We also quote the geometric parameters of the ANUBIS shaft configuration, for completeness.}
\label{tab:auxgeo}
\end{table}

For auxiliary detectors at the (HL-)LHC, the geometric parameters describing their location and size and the expected luminosity we used to derive our projected limits are listed in Table~\ref{tab:auxgeo}. We assume that LLPs will be reconstructed with an efficiency close to unity once they decay visibly inside the effective detector volume. For most detectors, the geometry and corresponding parameters are the same as in Ref.~\cite{Cheng:2024aco}, or slightly modified according to recent design updates. However, for ANUBIS a more significant update of the reach was performed, following the recent analysis of the ANUBIS collaboration~\cite{ANUBIS:2025sgg}. We choose the ceiling configuration as our benchmark, which is preferred experimentally at the current stage. The geometry of the detector volume is the gap between the outer radius of the ATLAS detector and the two-layered tracking station at the chamber ceiling. The angular coverage of the detector is about $\eta \in [-0.97,0.97]$ in pseudorapidity and $\phi \in [-1.1,1.2]$ in azimuthal angle. The exact geometry of the detector can be found in Ref.~\cite{ANUBIS:2025sgg}. Note that without extra shielding, the projected background level is significantly larger than order $1$ event at ANUBIS, in contrast with the other auxiliary detectors considered. The expected background at ANUBIS varies depending on the detector configuration and additional cuts. Using the ATLAS detector as an active veto can effectively suppress the background.
Requiring that the MET measured by ATLAS be greater than $30\;\mathrm{GeV}$, the total background for the ceiling configuration can be reduced to $\sim 180$ events~\cite{ANUBIS:2025sgg}, by scaling the background estimate of the ATLAS search for LLPs decaying to displaced hadronic jets in the muon spectrometer~\cite{ATLAS:2018tup}. After including the effect of systematic uncertainties, a new physics model which produces $36$ or more signal events will be excluded at 95\% CL. Based on simulation, we find that the extra MET requirement reduces the signal efficiency of
$Z$ decays to dark showers by a factor of $0.17$, almost independent of the dark meson masses and the $\he$ lifetime. This factor is included in our projections presented in Sec.~\ref{ssec:Zportal}.

\bibliography{DPDM}
\end{document}